\documentclass[%
 reprint,
nofootinbib,
 amsmath,amssymb,
 aps,
]{revtex4-2}

\usepackage{graphicx}
\usepackage[dvipsnames]{xcolor}
\usepackage{subcaption}
\usepackage{dcolumn}
\usepackage{bm}
\usepackage{xfrac}
\usepackage{hyperref}
\hypersetup{colorlinks=true,linkcolor=red,citecolor=blue,filecolor=blue,urlcolor=blue}
\usepackage{booktabs}
\usepackage{orcidlink}
\usepackage{physics}
\usepackage{multirow}
\usepackage{aas_macros}
\usepackage{url}
\usepackage{siunitx}
\usepackage{xspace}
\usepackage{comment}
\usepackage{textcomp}
\usepackage{tabularx}
\usepackage{calrsfs}

\DeclareMathAlphabet{\pazocal}{OMS}{zplm}{m}{n}

\newlength{\dhatheight}

\begin{document}

\title{
Characterization of a Two-Channel Optical and Near-infrared Transition Edge Sensor System for Rare-Event Searches
}

\author{Manuel Meyer \orcidlink{0000-0002-0738-7581}}
\email{mey@sdu.dk}
\affiliation{CP3-Origins, University of Southern Denmark, Campusvej 55, DK-5230 Odense M, Denmark}

\author{Katharina-Sophie Isleif}
\affiliation{Helmut-Schmidt-Universität, Holstenhofweg 85, 22043 Hamburg, Germany}

\author{Friederike Januschek}
\affiliation{Deutsches Elektronen-Synchrotron DESY, Notkestr. 85, 22607 Hamburg, Germany}

\author{Axel Lindner}
\affiliation{Deutsches Elektronen-Synchrotron DESY, Notkestr. 85, 22607 Hamburg, Germany}

\author{Gulden Othman}
\altaffiliation{Now at Munich Quantum Instruments GmbH,
Lichtenbergstraße~8, 85748 Garching, Germany
}
\affiliation{Helmut-Schmidt-Universität, Holstenhofweg 85, 22043 Hamburg, Germany}

\author{Elmeri Rivasto}
\affiliation{CP3-Origins, University of Southern Denmark, Campusvej 55, DK-5230 Odense M, Denmark}

\author{José Alejandro Rubiera Gimeno}
\affiliation{Helmut-Schmidt-Universität, Holstenhofweg 85, 22043 Hamburg, Germany}

\author{Christina Schwemmbauer}
\altaffiliation{Now at Munich Quantum Instruments GmbH,
Lichtenbergstraße~8, 85748 Garching, Germany
}
\affiliation{Deutsches Elektronen-Synchrotron DESY, Notkestr. 85, 22607 Hamburg, Germany}

\date{\today}

\begin{abstract}
Transition edge sensors (TESs) are superconducting energy‑resolving microcalorimeters that have demonstrated 
low background rates as well as
quantum efficiencies close to unity for photons at optical and near‑infrared wavelengths.
This makes these detectors well suited for rare‑event searches.  
We report on the comprehensive characterization of 
a two-channel detector module consisting of two tungsten TESs
optimized for the detection of photons with a wavelength of 1064\,nm. The devices achieve a system detection efficiency of $(86\pm1)\,\%$, an energy resolution better than 7\,\%, and a background dark-count rate
of photon-like events
below 6\,mHz 
when coupled to an optical fiber.
Using an unbinned likelihood framework, we find the dark count rate to be compatible with 
blackbody radiation from the room‑temperature laboratory environment. 
Thanks to the energy resolution of the TESs, we show that it is possible to  
detect monochromatic signals at 1064\,nm with 
photon rates $\geqslant 2.7_{-0.6}^{+0.8} \times10^{-5}\,\mathrm{Hz}$,
which corresponds to a power of $\geqslant (5.0_{-1.1}^{+1.4})\times10^{-24}\,\mathrm{W}$,
within 20\,days of measurement time at the $5\,\sigma$ confidence level.  
This makes our detectors well suited for searches for hypothetical axions and axion‑like particles with experiments such as the Any Light Particle Search~II (ALPS~II) or axion interferometers.  
The developed methodologies are not only applicable to axion searches, but are also relevant for rare‑event searches with TESs in general.
\end{abstract}

\graphicspath{{./}{./figures}}


\maketitle

\section{Introduction}
\label{sec:intro}

Transition edge sensors (TESs) are sensitive microcalorimeters operated in the region of the superconducting phase transition~\cite{2005cpd..book...63I,DeLucia:2024sxp,Fukuda:2024iwa}.
As a result, a temperature increase due to the absorption of electromagnetic radiation leads to a sharp increase of the detector's resistance. 
By integrating the TES in an electrical circuit, this change in resistance leads to a change in current which in turn can be read out through an inductively coupled superconducting quantum interference
device (SQUID).
These sensors are widely used as cryogenic single-photon detectors from sub-mm wavelengths to gamma-ray energies~\cite{2022JLwT...40.7578L}.
The combination of their high quantum efficiency, dead times of the order of microseconds, as well as energy resolution capability~\cite{2008OExpr..16.3032L,2022SuScT..35i5002H}, makes them ideal devices for a large range of applications including photonic quantum computing~\cite{2021Natur.591...54A,2022Natur.606...75M,2023NIMPA105468408L}, imaging in biophysics~\cite{2017NatSR...745660N,Niwa2021} and astrophysics~\cite{1999ApJ...521L.153R,2015SuScT..28h4003U}, as well as rare-event searches~\cite{2024EPJC...84.1001A,2024ApPhL.125w2601R}. 

In particular, TESs have been proposed as single-photon detectors in light-shining-through-a-wall (LSW) experiments~\cite{2013JInst...8.9001B,2015JMOp...62.1132D} and axion interferometers~\cite{2024PhRvD.109i5042Y} that 
are sensitive to a wide variety of light beyond-the-standard-model particles, collectively referred to weakly interacting slim particles (WISPs; see Ref.~\cite{Arza:2026rsl} for a recent review). 
Prominent examples of pseudoscalar WISPs are 
axions and axion-like particles (ALPs).
Axions are a consequence of the solution to the strong CP problem of QCD~\cite{1977PhRvD..16.1791P, 1978PhRvL..40..223W, 1978PhRvL..40..279W} and
both axions and ALPs could explain the observed density of cold dark matter~\cite{1983PhLB..120..133A, 1983PhLB..120..137D, 2012JCAP...06..013A}.
These pseudoscalar WISPs
are predicted to couple to photons in the presence of electromagnetic fields; this interaction is described through the Lagrangian density $\pazocal{L}_{a\gamma} = g_{a\gamma} \mathbf{E}\cdot\mathbf{B} a$, where $g_{a\gamma}$ is the photon-WISP coupling strength, $\mathbf{E}$ and $\mathbf{B}$ are the electromagnetic fields, and $a$ is the WISP field strength. 

Axion interferometers and the latest-generation LSW experiment called Any Light Particle Search~II (ALPS~II) employ high-power lasers at a wavelength of 1064\,nm together with high-finesse Fabry-P\'erot optical cavities to enhance a potential signal.
Once completed, ALPS~II is designed to feature two cavities each immersed in a strong magnetic field. 
Laser light oscillating in one cavity (the production cavity, PC) could convert into WISPs, which then leave the cavity, traverse a light-tight barrier, and reconvert back into photons in a second cavity (the regeneration cavity, RC).
ALPS~II recently finished its first science campaign 
using a heterodyne detection system, a single cavity (the RC), and two strings of 12 superconducting dipole magnets that provide a product of the magnetic flux density and length of the field that is over $568\,\mathrm{T}\cdot\mathrm{m}$ before and after the wall~\cite{2025OExpr..3311153K,2026arXiv260118684S}.
Axion interferometers, on the other hand, do not require a magnetic field but instead assume axions make up part if not all of the local dark matter density~\cite{2018PhRvD..98c5021D,2018PhRvL.121p1301O,2019PhRvD.100b3548L,2020PhRvD.101i5034M}. 
In this case, a p-polarized laser pump field propagating in a cavity could experience a polarization rotation inducing s-polarized light shifted in frequency by an amount proportional to the WISP mass. 
If the pump field can be efficiently suppressed by a series of polarizing beam splitters and cavities, single-photon detectors could detect the s-polarized component (or vice versa)~\cite{2024PhRvD.109i5042Y}. 
Such experiments require background rates $\lesssim 10^{-5}\,\mathrm{Hz}$.
This is a challenge for cryogenic single-photon detectors like a TES  
given the irreducible background of blackbody photons produced in the room-temperature laboratory environment that inevitably couples to the optical fiber guiding the signal to the TES~\cite{Rosenberg:2005zme}.

In this article, we present a comprehensive characterization
of a two-channel TES module intended as possible future single-photon detector at ALPS~II.
We describe the module in in Sec.~\ref{sec:tes} and present 
the measurement of the system detection efficiency in Sec.~\ref{sec:sde}. 
In Sec.~\ref{sec:calibration}, we discuss the energy calibration of our TESs with lasers at two different wavelengths and then introduce a method for selecting photon-like events to derive the detectors' energy resolution and full detector response matrix.
In Sec.~\ref{sec:bkg}, we develop an unbinned likelihood framework that allows us to use the full spectro-temporal information of background measurements to fit a blackbody background model to the photon data. 
With the derived background model, we assess the sensitivity of our TESs to detect a spectral line as expected in ALPS~II and axion interferometers in Sec.~\ref{sec:sensitivity}. 
The presented techniques are not limited to axion searches, but are applicable to rare-event searches with TESs in general.
We conclude and provide an outlook in Sec.~\ref{sec:conclusion}.

\section{The ALPS~II TES module }
\label{sec:tes}

Our detector module consists of two tungsten TES chips (called TES1 and TES2 hereafter) fabricated at the National Institute of Standards and Technology (NIST) and incorporated into an optical stack to maximize the absorption for near infrared photons at 1064\,nm~\cite{2008OExpr..16.3032L}.
Both TESs have a surface area of 25\,µm\,$\times$\,25\,µm, a thickness of 20\,nm, and a critical temperature around 140\,mK.
They are operated in voltage-biased condition at a working point (WP) corresponding to 30\,\% of the normal conducting resistance. 
We observe that TES2 has a stronger response to the applied bias current  
and it reaches the normal conducting state at a critical current $I_\mathrm{c} \sim 40\,$µA,  whereas the critical current for TES1 is twice as large. 
As a consequence, the applied bias current to TES2 to set the WP is lower (11.3\,µA) compared to the current applied to TES1 (33.8\,µA).

The energy deposited in the TESs is measured as a voltage output from a single-stage SQUID with 5\,GHz gain bandwidth product provided by Physikalisch-Technische Bundesanstalt (PTB), Berlin, Germany, and electronics from Magnicon. 
The SQUIDs are connected to the TESs via an intermittent copper pad that acts as a heat sink to reduce heating of TESs. 
The module with both TESs is situated inside a Bluefors~SD dilution refrigerator at 25\,mK base temperature. 
Single-mode optical fibers connect light sources at room temperature to the TESs. The connection between the fiber and TESs is realized with a zirconia fiber sleeve leading to minimal coupling losses~\cite{2011OExpr..19.9102M}.
Further details about the experimental setup can be found in Refs.~\cite{2022JLTP..tmp...93S, RubieraGimeno:2025bhr, Rivasto:2025gjk}. 



\section{System detection efficiency}
\label{sec:sde}

The system detection efficiency (SDE), $\eta$, is given by the ratio between power measured with the TES, $P_
\mathrm{TES}$, and the power sent to the TES, $P_{\mathrm{TES,~in}}$,
\begin{equation}
    \eta = \frac{P_{\mathrm{TES}}}{P_{\mathrm{TES,\,in}}} = n_\mathrm{TES}\frac{hc}{\lambda}\frac{1}{P_{\mathrm{TES,\,in}}}, 
    \label{eq:sde-base}
\end{equation}
where $h$ is Planck's constant, $c$ is the speed of light in vacuum, and $n_\mathrm{TES}$ is the photon rate measured with the TES. 
It is calculated by identifying and counting individual photon pulses, which we further describe in Sec.~\ref{sec:pulse-counting}. 
In our experimental setup (Sec.~\ref{sec:setup}), single-photon pulses could still be confidently identified without significant pile-up for powers $P_{\mathrm{TES,\,in}}\lesssim 1\,$fW. 
Measuring such low powers with commercially available photodiodes to calibrate our TES module is challenging. 
We overcome this issue by implementing a cascade of variable optical attenuators (VOAs), whose optical attenuation can be adjusted through an 
applied external
voltage. 
In the subsequent sections, we describe our experimental setup and how we can relate $P_{\mathrm{TES,\,in}}$ to measurements with our photodiodes.

\subsection{Experimental setup}
\label{sec:setup}
\begin{figure*}[htb]
    \centering
    \includegraphics[width=0.8\linewidth]{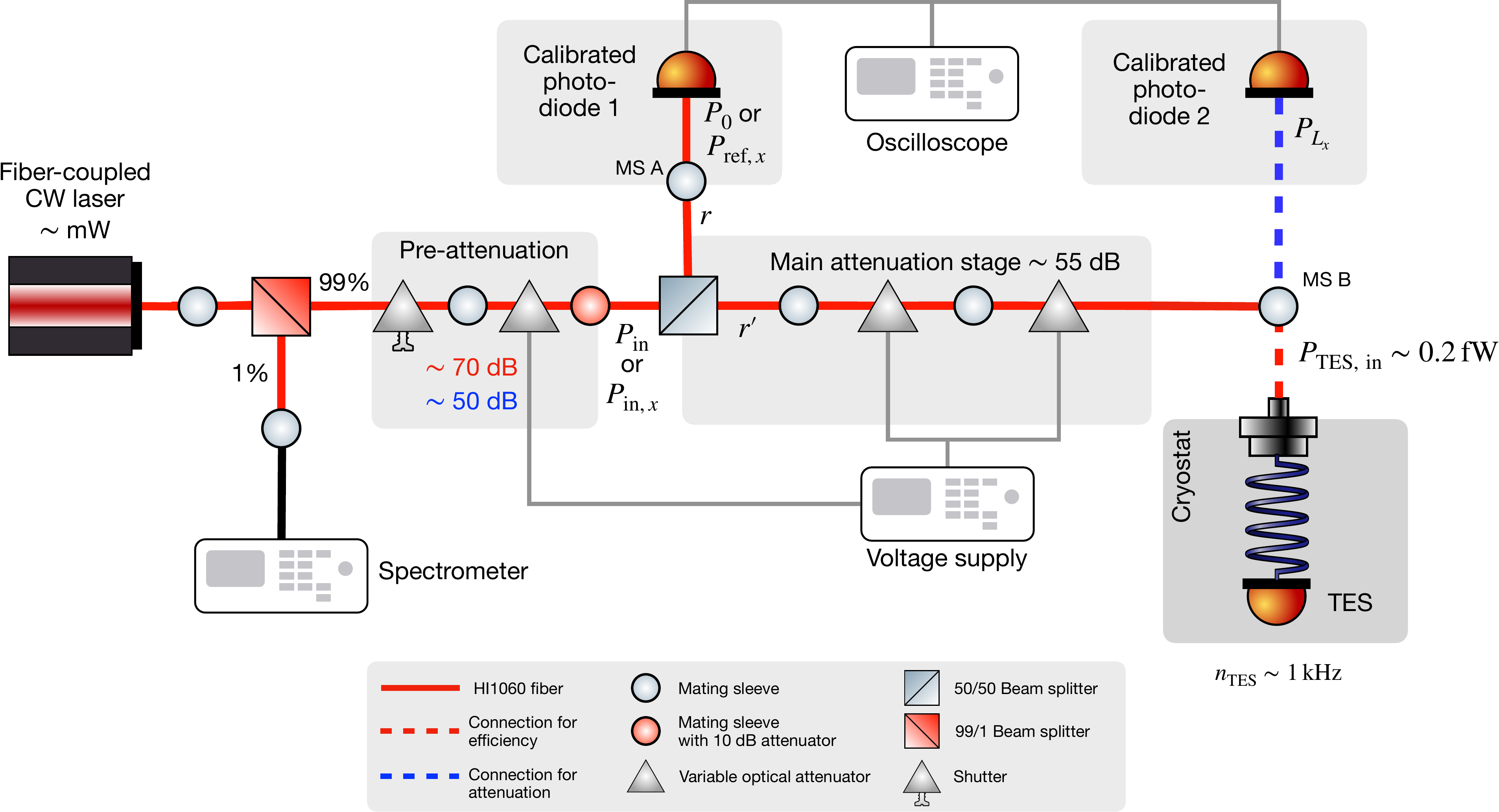}
    \caption{Experimental setup for the SDE measurement.
    The main components are the cw laser, two attenuation stages consisting of several optical attenuators, two PDs, and the TES. 
    Connections and values shown in blue refer to the attenuation calibration, connections and values in red show denote the setup for the SDE measurement.
    }
    \label{fig:eff-setup}
\end{figure*}

The experimental setup to measure the SDE is shown in Fig.~\ref{fig:eff-setup}. 
It is inspired by the setup reported in Ref.~\cite{Gerrits:2019idb}.
The main components are the detector module with two TES channels within the cryostat, two InGaAs photodiodes (PDs) 
with a wide dynamic range from picowatts to tens of nanowatts, as well as two adjustable attenuation stages. 
The noise-subtracted voltage output of the PDs is converted to optical power through a calibration (carried out at PTB) against known standards,
see further details in Appendix~\ref{sec:calpd}.

As illustrated in Fig.~\ref{fig:eff-setup}, a first beam splitter (BS), insertion losses, and the pre-attenuation stage reduce the optical power of a 
fiber-coupled continuous-wave (cw) 
laser with a nominal wavelength of 1064\,nm
from $\lesssim\,$mW by $\sim$70\,dB to $\sim$100\,pW during an SDE measurement.
The laser wavelength is monitored with a spectrometer (either an Ocean Insight HR4Pro or Ocean Insight NIRQuest) connected to the first BS with a fiber with thermally expanded core. 
The pre-attenuation is mainly achieved through one VOA, 
denoted as VOA0, with constant voltage supply and a fixed-value attenuator (providing $\sim$10\,dB attenuation at 1064\,nm). 
The pre-attenuation stage is also equipped with a second VOA whose attenuation can be adjusted manually with a screw.
This attenuator acts as a shutter to determine the noise of the PDs, $\bar{V}_\mathrm{noise}$ (screw fully closed, which provides an attenuation of $\sim$50\,dB).
We call the optical power transmitted in the fiber after the pre-attenuation $P_\mathrm{in}$.

A second BS with roughly equal split ratios $r'\sim r \sim 0.5$ splits $P_\mathrm{in}$ into $P_0 =r P_\mathrm{in} A_\mathrm{MS\,A}$, where $A_\mathrm{MS\,A}$ is the insertion loss of mating sleeve (MS)~A, and the power $r' P_\mathrm{in}$ to be attenuated further. The value of $P_0$ is monitored with PD1. 
It is related to the power sent to the TES module through
\begin{equation}
    P_\mathrm{TES,\,in} = r' A_\mathrm{tot} A_\mathrm{MS\,B} P_\mathrm{in}  = \frac{r'}{r}\frac{ A_\mathrm{MS\,B}}{A_\mathrm{MS\,A}} A_\mathrm{tot} P_0 ,\label{eq:p-tes-in}
\end{equation}
where $A_\mathrm{MS\,B}$ is the insertion loss at MS\,B and $A_\mathrm{tot}$ is the total attenuation provided by the main attenuation stage.
This stage consists of two additional VOAs, VOA1 and VOA2, with voltage supplies $V_1, V_2$ that are set during an SDE measurement to $V_1 = V_2 = 3.7\,\mathrm{V}$ such that a total attenuation of roughly $\sim55\,$dB is achieved, bringing down the optical power to acceptable levels for the TES.
Single-mode fibers with high transmission at 1064\,nm are used throughout the setup 
along with polarization-maintaining MSs and FC/APC fiber connectors. 

To calibrate $A_\mathrm{tot}$, a second PD (PD2) is used that can be connected to the second attenuation stage in lieu of the TES at MS\,B (blue connection in Fig.~\ref{fig:eff-setup}). 
As we show in Appendix~\ref{app:atten}, we can rewrite Eq.~\eqref{eq:sde-base} as
\begin{equation}
    \eta = n_\mathrm{TES}\frac{hc}{\lambda} \frac{1}{A P_0},
\label{eq:sde-final}
\end{equation}
where $A$ is the attenuation of VOA1 and VOA2 including their insertion losses and losses at the intermittent MSs, see Eq.~\eqref{eq:atten-final}. 

During the same cool down, we conducted two measurement campaigns of the SDE 4\,days apart for both TES channels,  
which we refer to SDE-A and SDE-B in the following. 
To estimate the uncertainty introduced by breaking the fiber connection at MS\,B when connecting PD2, we dis- and reconnect the optical fiber at MS\,B multiple times.
This was done each time after taking time traces with the different TES channels and after each round of measuring the attenuation $A$. 
When replugging the fiber on MS\,B for the attenuation measurements, we inserted the FC/APC connector for PD2 such that the power was maximized on the oscilloscope screen. 
During the second measurement,
we also performed a second measurement round in which we simply screwed-in the FC/APC connector as tight as possible, leading to slightly lower optical powers. 
We will refer to this set of measurements to SDE-B-tight.

Before turning to the results, we briefly describe how we identify photon pulses in the measured time traces in the following section. 

\subsection{Counting Photon Pulses}
\label{sec:pulse-counting}

\begin{figure*}[htb]
    \centering
    \includegraphics[width=0.7\linewidth]{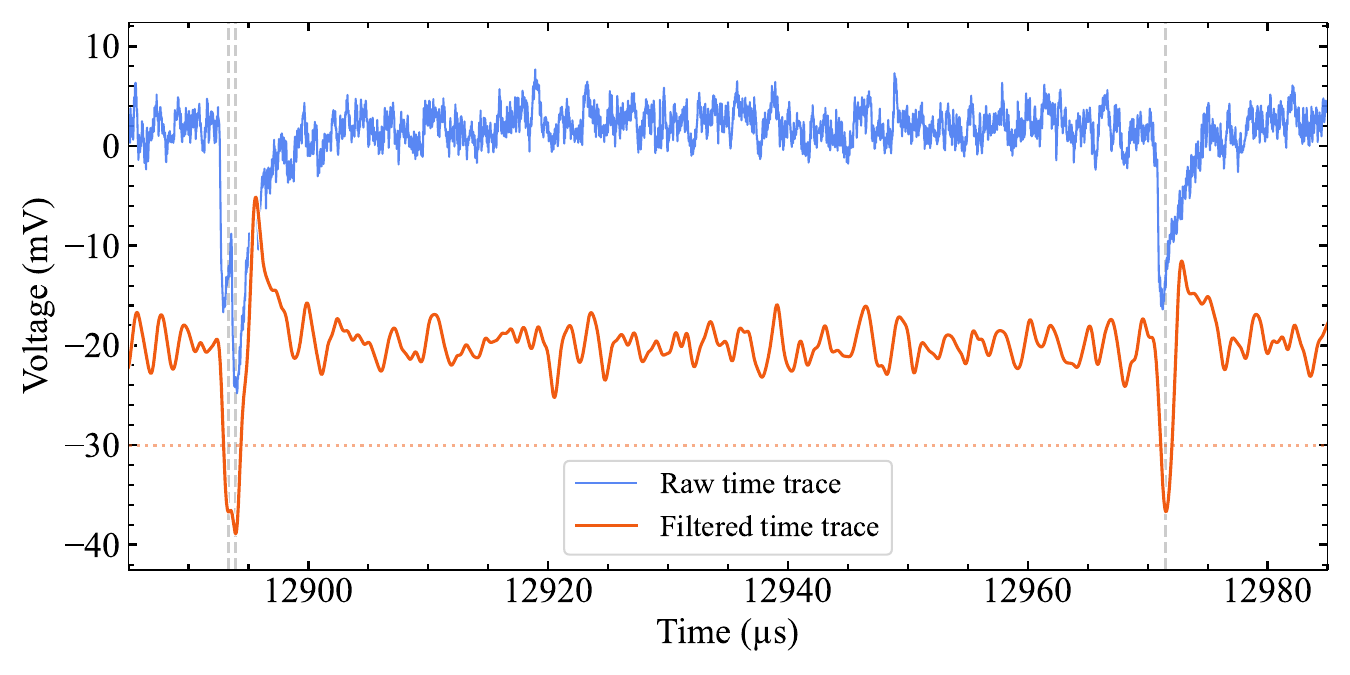}
    \caption{100\,µs example of a 4\,s measurement of the heavily attenuated light source with TES1. Three photon pulses are visible, where the first two pulses overlap. The filtered trace is shifted by $-20\,$mV for better visibility together with the minimum threshold (orange dotted line) required for peak detection. The found peaks are indicated by grey vertical dashed lines.}
    \label{fig:detected-pulse}
\end{figure*}

In each SDE measurement, we record 5 time traces with both TES channels.
Each time trace has a total length of 4\,s which is digitized with an ADC (AlazarTech ATS9626 16-bit digitizer) at a sampling rate of 50\,MHz resulting in $2\times10^8$ recorded samples. 
As an example, we show 100\,µs of such a trace in Fig.~\ref{fig:detected-pulse}. 
Note that with the applied attenuation, we expect optical powers $P_\mathrm{TES,\,in}
\sim0.2\,\mathrm{fW}$, which corresponds to a photon rate of $\sim 1\,\mathrm{kHz}$ for 1064\,nm photons.

For pulse finding, we apply two digital filters in post-processing: a third-order Butterworth low-pass filter with a cut-off frequency of 1\,MHz followed by an asymmetric trapezoidal filter~\cite{1994NIMPA.345..337J} with raising, constant, and falling parts with lengths of 10, 50, and 20 samples, respectively. 
The filtered trace is also shown in Fig.~\ref{fig:detected-pulse}. 
Next, a peak finding algorithm is applied, where the threshold for peak detection is set to half the minimum of the recorded trace and the width of the peak is required to be at least 5 samples.\footnote{
The width is determined as the number of samples with at least half the derived peak prominence, i.e., how much a peak stands out against its surroundings. See the \url{https://docs.scipy.org/doc/scipy/reference/generated/scipy.signal.find_peaks.html} for further details. 
}
Our algorithm reliably identifies single-photon peaks even in the presence of pile-up, as long as the time interval between the arrival of two photons is  $\Delta t \gtrsim 0.5\,$µs, as evident from the first two pulses in Fig.~\ref{fig:detected-pulse}. 

\begin{figure}[htb]
    \centering
    \includegraphics[width=0.85\linewidth]{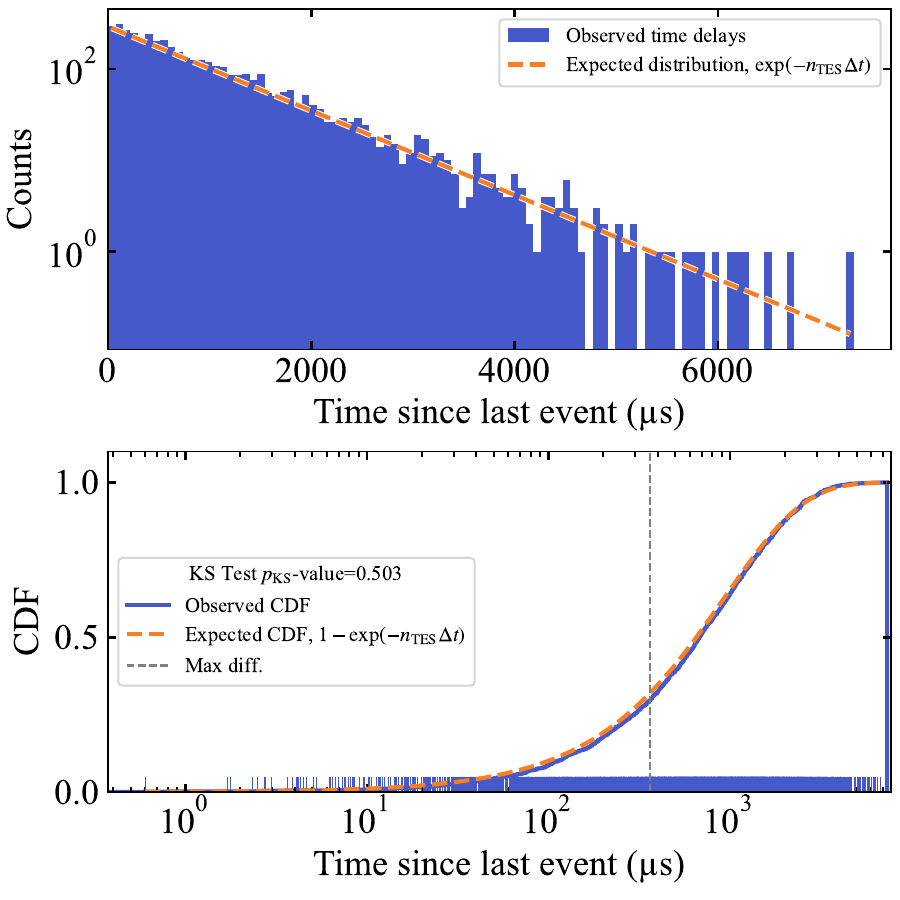}
    \caption{Distribution of time intervals between detected photon pulses for one measurement with TES1. \textit{Top:} observed distribution. \textit{Bottom:} Cumulative distribution. Individual events are shown as ticks along the $x$-axis. The time of maximum difference that enters the KS test is shown as a grey dashed line.}
    \label{fig:arrival-times}
\end{figure}

To 
ensure that our algorithm is suitable for reliable pulse identification at the given attenuation of the cw laser source, we test whether the time intervals between photons follow the expectation from Poisson-distributed events. 
For such events, the probability distribution $p(\Delta t; n_\mathrm{TES})$ of time intervals $\Delta t$ between consecutive pulses at an expected photon rate $n_\mathrm{TES}$ should follow an exponential distribution, $p(\Delta t; n_\mathrm{TES}) \propto \exp(-n_\mathrm{TES} \Delta t)$. 
The expected rate is simply derived from the number of identified pulses divided by the total length of the trace (4\,s). 
We plot both the observed distribution of $\Delta t$ and its cumulative distribution function (CDF) in the top and bottom panels of Fig.~\ref{fig:arrival-times}, respectively.

The observed distributions nicely follow the exponential expectation. 
This is also confirmed quantitatively by comparing the observed and expected CDFs by means of a Kolmogorov-Smirnov (KS) test. 
The probability that the observed distribution is drawn from the expected one for the example in Fig.~\ref{fig:arrival-times} is $p_\mathrm{KS} = 0.503$. 
For all conducted photon counting measurements, we find a mean of $\langle p_\mathrm{KS}\rangle = 0.574$ and standard deviation $s_\mathrm{KS} =  0.260$. The lowest $p_\mathrm{KS}$ value is 0.075 (see Tab.~\ref{tab:eff-measurements} in Appendix~\ref{app:results-errors} for all results).\footnote{
The minimum value found for $p_\mathrm{KS}$ is fully compatible with the probability to find at least one $p_\mathrm{KS} \leqslant \min(p_\mathrm{KS})$ in 20 independent trials conducted here; this probability is given by $1 - (1-\min(p_\mathrm{KS}))^{20}\approx0.79$. 
} 
We therefore conclude that the chosen attenuation does not saturate our TESs and that we reliably identify photon pulses in the measured traces.

\subsection{Results}
\label{sec:eff-results}

\begin{figure*}
    \centering
    \includegraphics[width=0.8\linewidth]{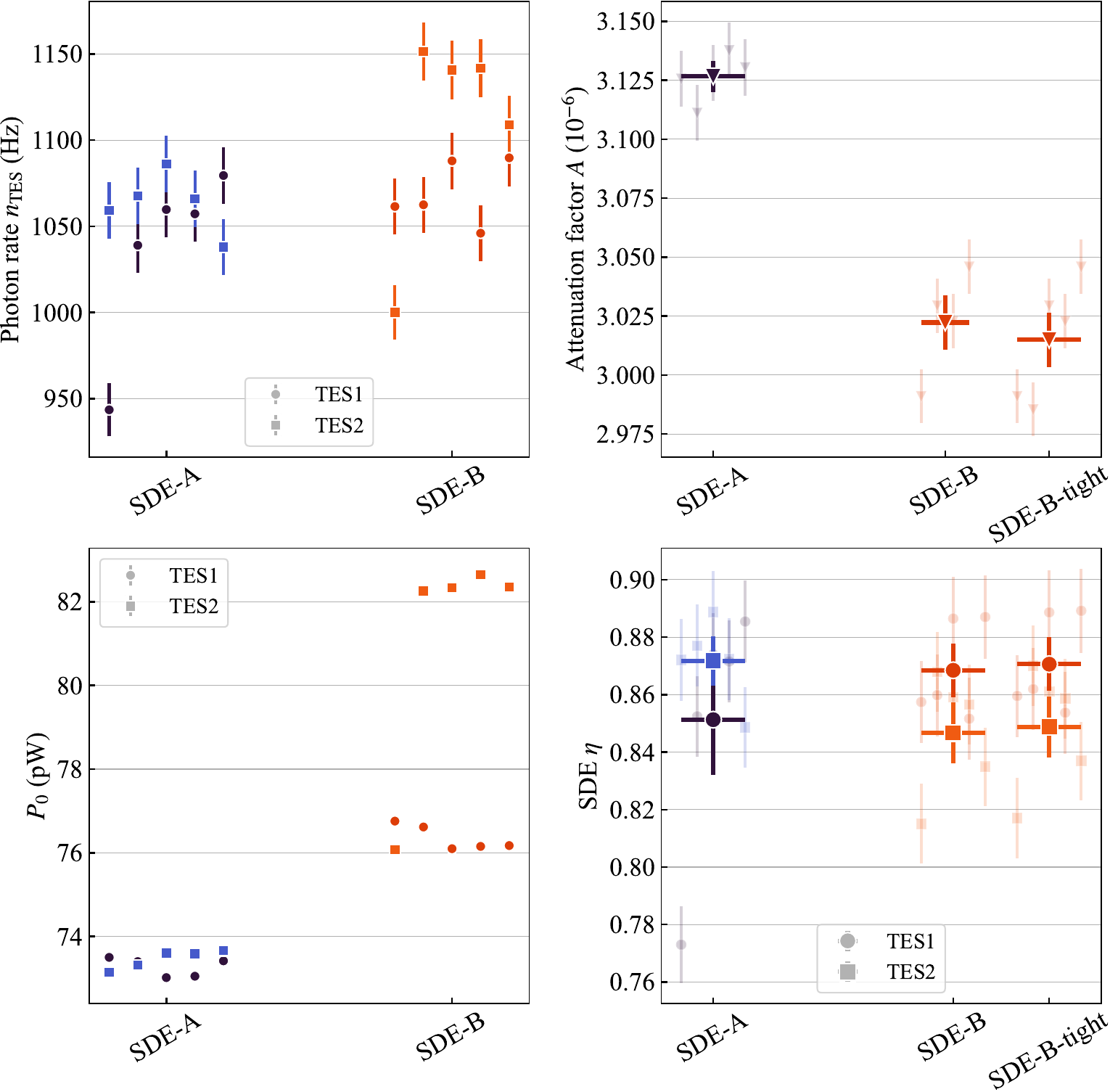}
    \caption{Results of the SDE measurements. Markers indicate which TES was taking data during the  $n_\mathrm{TES}$ and $P_0$ measurements. 
    Different color palettes (purple and dark-orange) indicate the two measurement campaigns SDE-A and SDE-B.
    \textit{Top left:} the recorded photon rates. Between each measurement, the fiber leading to TES cryostat is reconnected with MS~B. \textit{Top right:} The individual and averaged attenuation factors. Between each data point, PD2 was reconnected with MS~B.
    \textit{Bottom left:} The optical powers measured at PD1 during the corresponding photon rate measurements with the TES channels. 
    \textit{Bottom right:} Final derived SDE values using the corresponding average attenuation factors from the top right panel. The uncertainty for TES1 for SDE-A is twice as large as the uncertainties for the other dates and other channel due to the first outlier measurement as explained in Sec.~\ref{sec:eff-results} . 
    }
    \label{fig:eff-results}
\end{figure*}

The results of our two measurement campaigns for both TES channels are shown in the four panels of Fig.~\ref{fig:eff-results} 
including the main result of the SDE derived with Eq.~\eqref{eq:sde-final} in the bottom right panel.
The left panels show the measured photon rate $n_\mathrm{TES}$ (top) and the simultaneously measured optical power $P_0$ incident on PD1 (bottom).   
For most measurements, we observe a photon rate around 1.05~kHz with uncertainties derived from Poisson noise on the counted number of photons (see Appendix~\ref{app:results-errors} for a detailed discussion of the uncertainties).  
Notable exceptions of the photon rate are the very first measurement with TES1 for SDE-A as well as the last four measurements with TES2 for SDE-B. 
The first measurement with TES1 is clearly an outlier: the photon rate is significantly lower by a factor 0.89 compared to the average rate of the next 4 measurements. 
This can not be explained by Poisson noise or power fluctuations of the laser source since $P_0$ is relatively constant over all 5 measurements. 
As a result, the derived efficiency for this measurement is also significantly lower at 77\,\% (lower left panel of Fig.~\ref{fig:eff-results}) compared to the other 4 measurements where the efficiency is around $87\,\%$.
We believe that the most likely explanation for this result is a sub-optimal fiber connection at MS~B during this measurement.
The last four measurements with TES2 during SDE-B show a significantly higher photon rate compared to the first measurement with TES2 and the five measurements with TES1 performed just before these last 4 measurements. 
The increased photon rate is due to an increase of optical power, which is clearly visible in $P_0$.
This is either due to a fluctuation of laser power or in the voltage supply for VOA0 in the pre-attenuation stage. 
The results for all individual measurements of $n_\mathrm{TES}$ and $P_0$ are reported in Tab.~\ref{tab:eff-measurements} in Appendix~\ref{app:results-errors}.

With the shutter in the pre-attenuation stage closed, we took additional measurements with both TES modules to evaluate whether dark counts could affect the photon rate. We took 100\,s of data and recorded photon pulses exceeding a trigger threshold of 10\,mV. This resulted in a dark current rate of $\sim0.1\,$counts per second. Thus, the effect on the final efficiency is completely negligible.

The measured attenuation factors $A$ are shown in the top right panel of Fig.~\ref{fig:eff-results}. 
Individual measurements are shown in light colors with the average values for each round of measurements plotted on top.
The exact values are also reported in Tab.~\ref{tab:atten-factor} in Appendix~\ref{app:results-errors}.
The average attenuation for both dates is around $55\,$dB.
Between the two dates, all fibers where disconnected, the fiber tips and MSs cleaned, and the fibers reconnected.
The average attenuation factor $\langle A \rangle$ was about a factor of 1.03 larger for SDE-A, which can be explained by slightly different insertion losses at the MSs of the attenuation stage (cf. Fig.~\ref{fig:eff-setup}).  
As explained in Sec.~\ref{sec:setup}, during the second measurement campaign we performed two separate measurements where we adopted different techniques to connect the optical fiber leading to PD2 with MS\,B. 
The resulting difference between the two average values is $\sim 0.2\,\%$, which makes us confident that overall our results on the attenuation factor are not dominated by the exact way how we connect the fibers at the mating sleeves.

The final SDE results are shown in the bottom left panel. 
The SDE is derived from Eq.~\eqref{eq:sde-final} where we use the average value of the attenuation factor $\langle A \rangle$ for each measurement day. 
The average values obtained for the SDE are reported in Tab.~\ref{tab:eff-results}.
The measurements at different dates are compatible with each other within uncertainties. Both TES modules show similar performance with an average SDE $\langle \eta \rangle \gtrsim 86\,\%$ with uncertainties of the order of $1\,\%$. 
The major contributor to the overall uncertainty is the Poisson noise from the counted number of photons.
The fact that we observe comparable SDE values for both detectors and only a small scatter between the single SDE measurements makes us confident that our experimental setup provides a reliable measurement of the SDE.
Although our SDE values are below the record efficiencies achieved with W and Ti-based TESs of 95\,\% and 98\,\%, respectively \cite{2008OExpr..16.3032L, 2011OExpr..19..870F},
they agree well with other SDE values reported in the literature measured typically at 850\,nm or 1550\,nm \cite[e.g.,][]{Schmidt2018,2022JLwT...40.7578L,2025arXiv250208952E}. 

\begin{table}[htb]
\caption{\label{tab:eff-results} Average SDE values obtained for the two measurement dates. For the second date, two different protocols are used to tighten the fiber-MS connection at MS~B.
The final average values are obtained from averaging the measurements SDE-A and SDE-B.
}
\centering
\begin{tabular}{c|c}
\hline
\hline
Sensor & Average SDE $\langle \eta \rangle$ \\
\hline
\multicolumn{2}{c}{SDE-A}\\
\hline
TES1 & 0.851 $\pm$ 0.019 \\
TES2 & 0.872 $\pm$ 0.008 \\
\hline
\multicolumn{2}{c}{SDE-B}\\
\hline
TES1 & 0.871 $\pm$ 0.009 \\
TES2 & 0.849 $\pm$ 0.010 \\
\hline
\multicolumn{2}{c}{SDE-B-tight}\\
\hline
TES1 & 0.873 $\pm$ 0.009 \\
TES2 & 0.852 $\pm$ 0.010 \\
\hline
\hline
\multicolumn{2}{c}{Final average values}\\
\hline
TES1 & 0.861 $\pm$ 0.011 \\
TES2 & 0.861 $\pm$ 0.007 \\
\hline
\end{tabular}
\end{table}

\section{Energy Calibration and Detector Response}
\label{sec:calibration}

The SDE alone is not sufficient to fully characterize the spectrum of a photon source;
we also require energy calibration and the energy resolution of the detector. 
These quantities should be derived from as clean as possible samples of photons at defined energies. Thus we also need to be able to efficiently reject background events.

To this end,
we carry out a pulse shape analysis, i.e., we fit a function describing the voltage response of the TES to the observed time traces.
The fit parameters are then used to discriminate photons from non-photon-like
background events and to estimate the deposited energy. 
Common functional forms are the TES response predicted by small signal theory (either fitted in time or frequency domain by Fourier transforming the signal and model~\cite{RubieraGimeno:2023}), or purely phenomenological functions. Following previous works~\cite{2022JLTP..tmp...93S}, we use an exponential rise and decay function, 
\begin{equation}
    V(t) = C - \frac{2V_0}{\exp\left(\frac{t_0 - t}{\tau_\mathrm{rise}}\right) + \exp\left(\frac{t - t_0}{\tau_\mathrm{decay}}\right)},
    \label{eq:pheno-func}
\end{equation}
with constant offset $C$, amplitude $V_0$, rise and decay times $\tau_\mathrm{rise}$ and $\tau_\mathrm{decay}$, respectively, and $t_0$ the onset of the pulse. 

We record a large number of pulses while the TES is connected to 
lasers emitting at different wavelengths. 
At the time of the measurement, two lasers were available: the laser for the SDE 
with central wavelength of $(1055.0\pm0.6)\,$nm 
and a red fiber checker laser with central wavelength of $\lambda=(654.9\pm0.2)\,$nm Both central wavelengths are determined with our HR4Pro Spectrometer
(see also Appendix~\ref{app:results-errors}).
It should be noted that the laser linewidths are much smaller than the energy resolution of the two TES channels, so that the laser photons can be regarded as monochromatic. 
The experimental setup is essentially identical to the one shown in Fig.~\ref{fig:eff-setup}, except for different lasers and attenuations used. 
We refer to these data as our calibration data sets and summarize their main characteristics in Tab.~\ref{tab:cal-dataset}.

\begin{table*}[]
    \centering
    \caption{Summary of the calibration data sets, likelihood analysis thresholds, and efficiencies to determine whether events are classified as photon-like.
    Uncertainties on likelihood thresholds and fraction of photon-like traces are derived from $k$-fold cross validation.
    In contrast to the SDE measurement, we record 200\,µs time lines at a sampling rate of 50\,MHz when the voltage exceeds a trigger threshold value of 12\,mV (22\,mV and 20\,mV) for the 
    1055\,nm
    (655\,nm) laser. 
    The traces contain $10^4$ samples, where 1{,}500 samples are recorded before the trigger time.
    }
    \label{tab:cal-dataset}
    \begin{tabular}{c|cc|cc}
    \hline
    \hline
      &  \multicolumn{2}{c|}{TES1} &  \multicolumn{2}{c}{TES2}\\
     \hline
     Central wavelength (nm) & $1055.0\pm0.6$ & $654.9\pm0.2$ & $1055.0\pm0.6$ & $654.9\pm0.2$ \\
     Central energy (eV) & $1.175\pm0.001$ & $1.893\pm0.001$ & $1.175\pm0.001$ & $1.893\pm0.001$ \\
     \hline
     Total measurement time (s)  & 20  & 20 & 30 & 20\\
     Trigger threshold (mV) & $-12$ & $-22$ & $-12$ & $-20$ \\
     Number of recorded pulses & 6053 & 5029 & 9651 & 4859 \\
     \hline
     Likelihood threshold $\xi$ & \multicolumn{2}{c}{$5.58 \pm  0.17$} & \multicolumn{2}{|c}{$-0.41 \pm  0.03$} \\
     Fraction of photon-like traces, test set & \multicolumn{2}{c}{$(98.97 \pm 0.15)$\,\%} & \multicolumn{2}{|c}{$(94.96 \pm 0.32)$\,\%} \\
     Fraction of photon-like traces, whole calibration set & \multicolumn{2}{|c}{\multirow{2}{*}{$(94.49 \pm 0.04)$\,\%}} & \multicolumn{2}{|c}{\multirow{2}{*}{$(91.00 \pm 0.06)$\,\%}} \\
     (before cleaning) & & & & \\
     \hline
     
    \end{tabular}
\end{table*}

For each recorded time line, we select a 30\,µs window with $N = 1{,}500$ sample containing a single pulse 
(with 250 samples before the trigger time)\footnote{We use the same algorithm for pulse identification as in Sec.~\ref{sec:pulse-counting}. If more than one pulse is present, we discard the trace as this could potentially bias the best-fit parameters.} 
from the full 200\,µs trigger window
and fit it with the phenomenological function of Eq.~\eqref{eq:pheno-func} by minimizing the $\chi^2$,
\begin{equation}
\chi^2 = \sum_{i = 1}^N \left(\frac{V_{i} - V(t_{i})}{\sigma_{i}}\right)^2,
\end{equation} 
where the uncertainty $\sigma_{i}$ of each voltage reading $V_{i}$ at time $t_{i}$ is estimated from the average of the main diagonal of the noise covariance matrix.
The covariance matrix is estimated from all recorded pulses using the last 1{,}500  samples of each full trace. 
The (subdominant) off-diagonal elements are neglected due to computational efficiency. 
We perform the $\chi^2$ minimization numerically using \texttt{Minuit}~\cite{James:1975dr,dembinski_2025_17565861}. 
The pulses are generally well described by our phenomenological function;
the resulting reduced $\chi^2$ values are typically below or close to unity.\footnote{ 
Reduced $\chi^2$ values below one can be explained by the fact the we neglected the off-diagonal contributions of the covariance matrix in the fit.
}
Thus, for each pulse, we are left with the best-fit parameters $\boldsymbol{\theta} = (C, V_0, \tau_\mathrm{rise}, \tau_\mathrm{decay}, t_0)$ along with the $\chi^2$ value of the fit. 
In the subsequent sections, we will use these parameters to select photon-like events (Sec.~\ref{sec:photonness}) and estimate the energy of these events (Sec.~\ref{sec:energy_calibration}), 
before constructing the full detector response matrix (Sec.~\ref{sec:drm}).

\subsection{Selection of photon-like events}
\label{sec:photonness}

Pulse shape analysis can be used to discriminate photon-like against background events caused by, e.g., non-photonic energy depositions in the TES or the surrounding substrate. 
Such non-photon-like events can be caused by, e.g., radioactive decays as well as cosmic rays and their secondaries.
They are usually of comparably high energy and the resulting traces can typically be easily distinguished from photon-like events as demonstrated previously~\cite{2022JLTP..tmp...93S, Meyer:2023ffd,Rivasto:2025gjk}. 
Another source of pulses is electrical noise~\cite{Manenti:2024etv,Schwemmbauer:2025evp}.

One common approach is to fit the distributions of the parameters describing a pulse with normal distributions.
Pulses are then considered photon-like when their parameters are within a certain interval (measured in units of the standard deviations) around the mean of that distribution~\cite[see, e.g.,][]{2015JMOp...62.1132D}. 
These intervals can also be chosen to be asymmetric around the mean~\cite{2022JLTP..tmp...93S,RubieraGimeno:2025bhr}.
Alternatively, clustering algorithms of the best-fit parameters (or linear combination of the parameters derived form principal component analysis) can be employed to identify different sources of energy depositions~\cite{Manenti:2024etv}. 
Furthermore, machine-learning algorithms have shown promising results in discriminating photon-like from non-photonic events~\cite{Meyer:2023ffd}.

Here, we follow the approach of describing the distribution with normal distributions. 
Importantly, we want to classify photon-like events independent of the photon energy, which is crucial for determining the spectrum of the signal and the photonic background (discussed below in Sec.~\ref{sec:bkg}).
Therefore, we first combine the data sets taken at different wavelengths and perform minimal data cleaning to ensure a minimum contamination of the data set with background events or false triggers caused by electrical noise. 
For this, we reject traces for which (a) more than one pulse is present, (b) the $\chi^2$ minimization does not yield a valid minimum according to the criteria of the \texttt{Minuit} optimizer and (c) the pulses resulting in the highest 5\,\% quantile of the overall distribution of $\chi^2$ values. 
These cuts retain $\sim95\,\%$ of the original data. 
Next, we randomly split the data into a training and test set at a ratio of 4:1 and fit the distribution of the best-fit parameters and the $\chi^2$ value, $\boldsymbol{\theta}' = (\tau_\mathrm{rise}, \tau_\mathrm{decay}, t_0, C, \chi^2)$ of the training set with a 5-dimensional multivariate normal distribution $\mathcal{N}(\boldsymbol{\mu}, \boldsymbol{\Sigma})$ with mean vector $\boldsymbol{\mu}$ and covariance matrix $\boldsymbol{\Sigma}$.
We exclude the best-fit pulse amplitude $V_0$ as it is directly proportional to the deposited energy.

For each pulse in the training set, we calculate the negative log-likelihood under the multivariate Gaussian model, 
\begin{equation}
    -\ln\pazocal{L}_\gamma(\boldsymbol{\mu}, \boldsymbol{\Sigma} | \boldsymbol{\theta}') = \frac{1}{2}\left(\boldsymbol{\theta}' - \boldsymbol{\mu}\right)^T\boldsymbol{\Sigma}^{-1}\left(\boldsymbol{\theta}' - \boldsymbol{\mu}\right),
    \label{eq:lnl-gamma}
\end{equation}
where constant terms have been neglected.
Using the distribution of likelihoods for the training set, we define a threshold $\xi$ such that events with $-\ln\pazocal{L}_\gamma < \xi$ are classified as photon-like. 
Choosing lower values of $\xi$ will result in a more efficient suppression of background at the cost of a lower efficiency for photon events. 
As most traces in our training sets should be unique photon events, we set $\xi$ such that 95\,\% (90\,\%) of all events in the training set for TES1 (TES2) are retained. 
Using $k$-fold cross validation, we repeat this procedure for a total of 5 partitions of the data into training and test sets. 
The resulting mean values of $\xi$ are reported in Tab.~\ref{tab:cal-dataset} along with the efficiencies for the test set and all recorded traces prior to data cleaning. 
Compared to previous analyses, our likelihood-based approach has the advantage that only a single value ($\xi$) is tuned for photon selection instead of multiple parameters for each separate distribution of best-fit values. 

\begin{figure*}[htb]
\centering
\includegraphics[width=.48\linewidth]{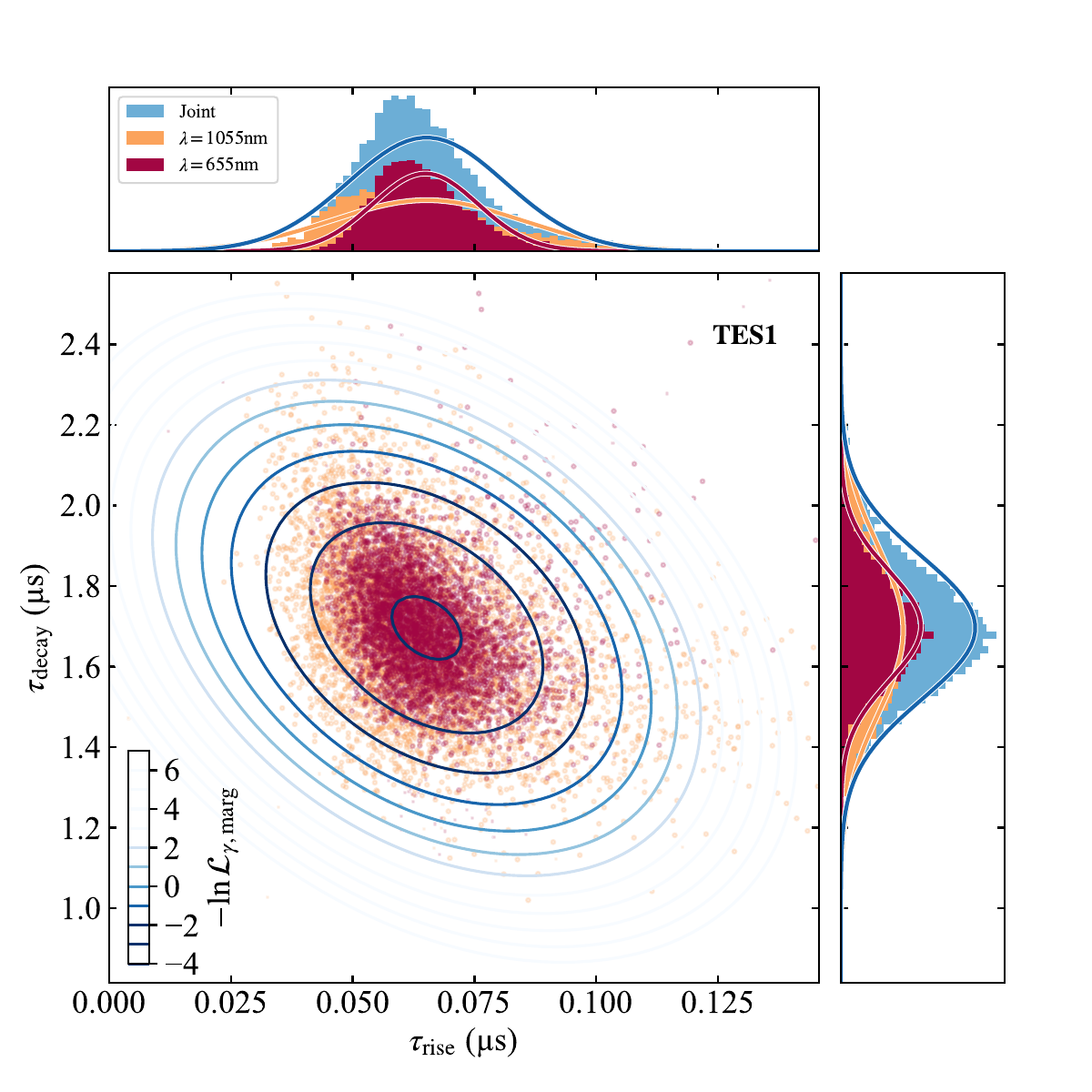}
\includegraphics[width=.48\linewidth]{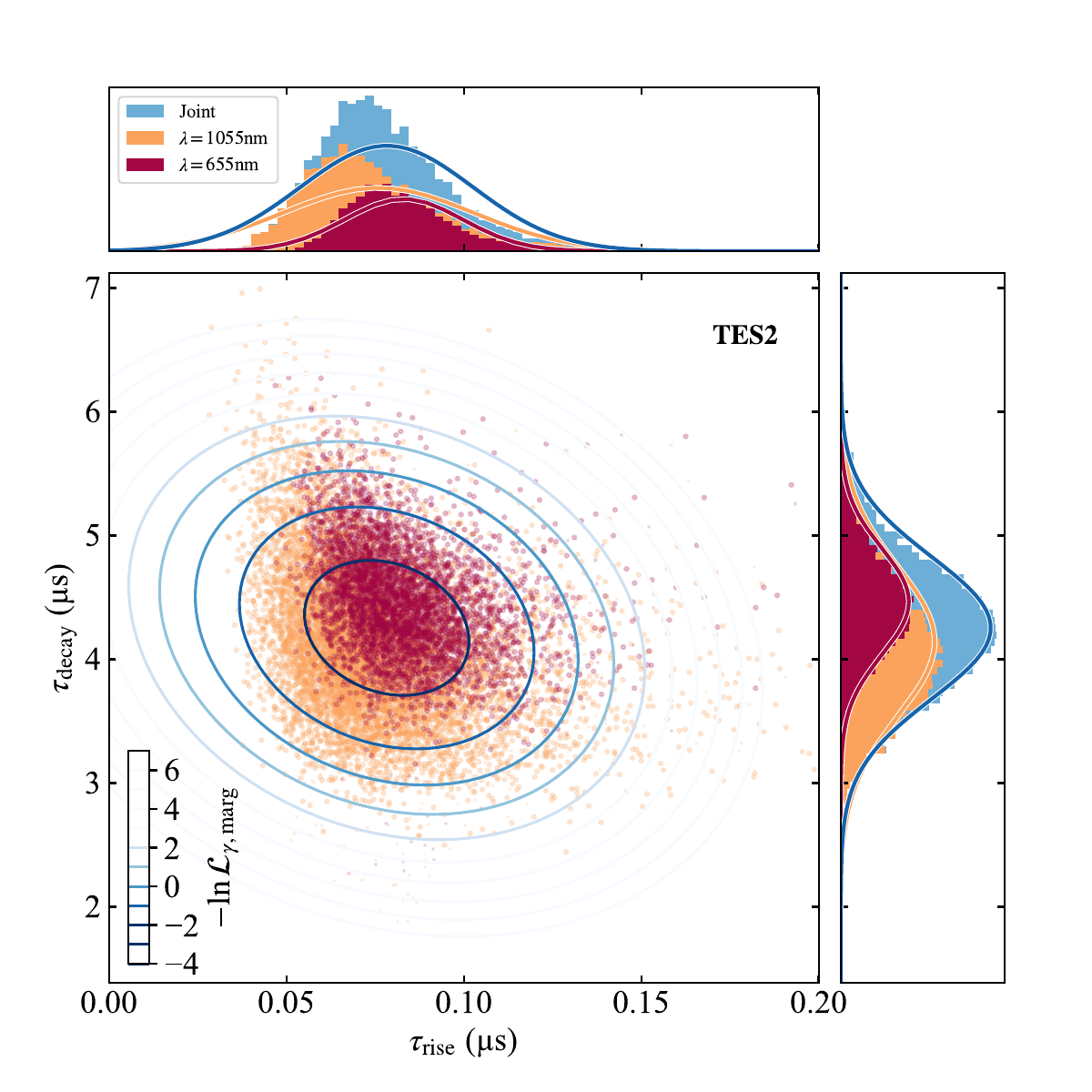}
\includegraphics[width=.48\linewidth]{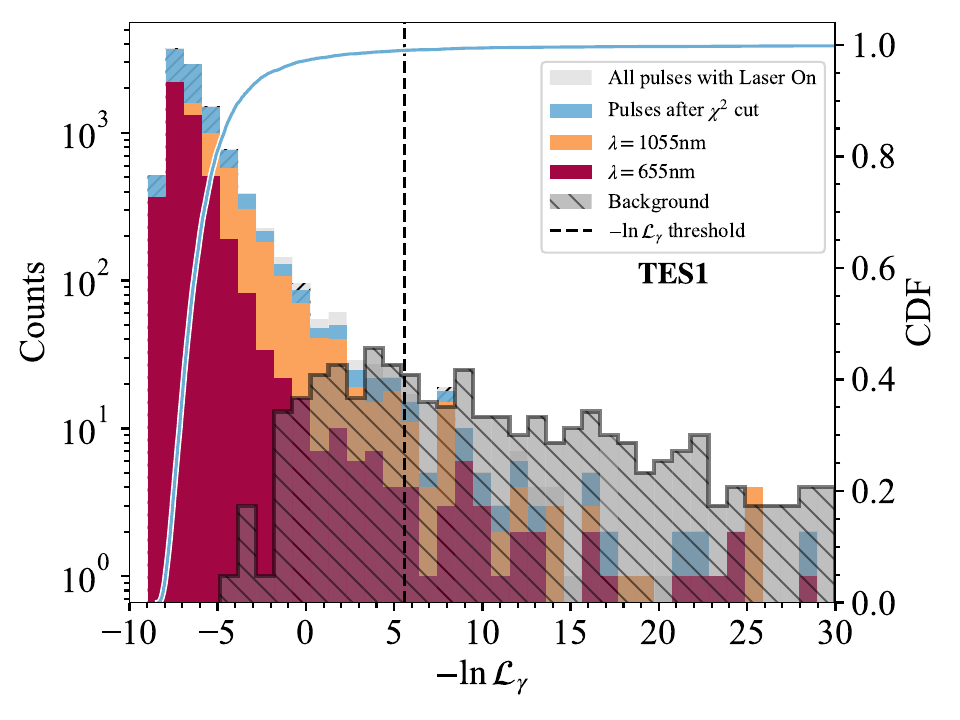}
\includegraphics[width=.48\linewidth]{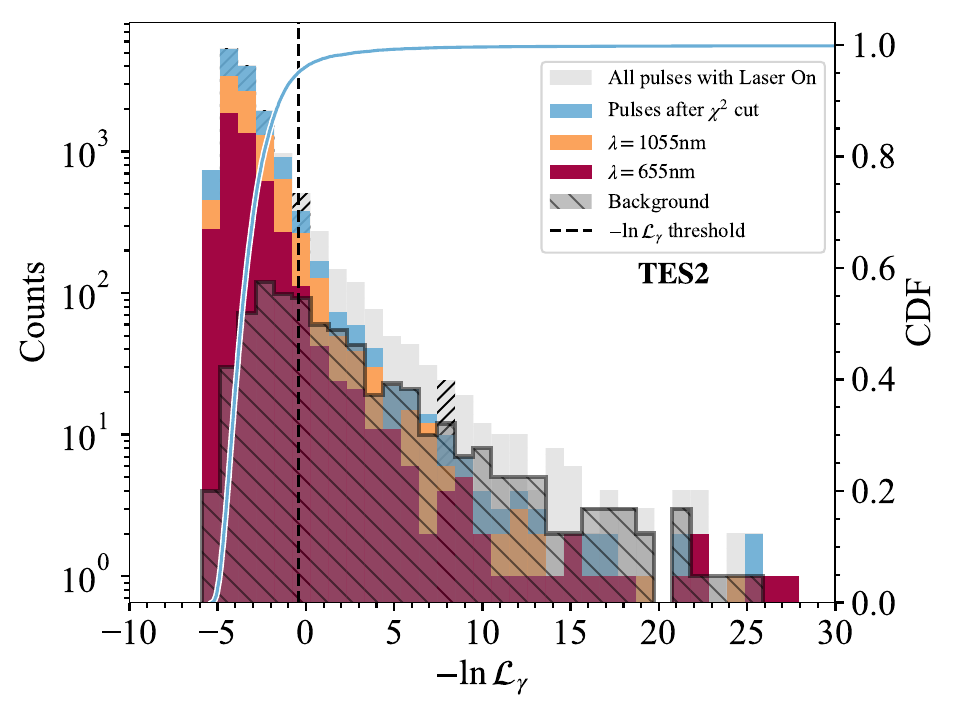}
\caption{\textit{Top:} distribution of best-fit parameters $\tau_\mathrm{rise}$ and $\tau_\mathrm{decay}$ for TES1 (left) and TES2 (right) for pulses recorded with lasers at two different wavelengths. 
The insets show the marginalized observed distributions along with the marginalized fitted normal distributions. 
The joint distributions when combining pulses at both wavelengths are also shown in insets.
Contours of constant values of the log-likelihood of the multivariate normal distribution marginalized over $t_0$, $C$, and $\chi^2$ are shown as solid lines. 
\textit{Bottom:} Negative log-likelihood values for the recorded pulses under the multivariate normal distribution derived from a training data set. The chosen threshold values of $\xi$ are shown as vertical dashed lines. The distribution of background data (discussed in Sec.~\ref{sec:bkg}) is shown as grey hatched histograms.
}
\label{fig:photonness}
\end{figure*}

We show the normal distribution marginalized over $t_0, C$ and $\chi^2$ in the top row of Fig.~\ref{fig:photonness} for both TES channels, together with contours of constant (marginalized) likelihood values.\footnote{Marginalizing a multivariate normal distribution results again in a normal distribution.} 
The insets show the one-dimensional normal distributions when all but one parameter (either $\tau_\mathrm{rise}$ or $\tau_\mathrm{decay}$) are marginalized over. 
For both channels and both wavelengths, one observes that the best-fit values of the decay time, $\tau_\mathrm{decay}$, are well described by normal distributions. 
We also note that the decay times for TES2 are in general more than a factor of two larger than for TES1.
The most likely reason for this is a higher low-frequency loop gain under constant current, $\mathcal{L}_I$, for TES1 due to a higher applied bias current at the WP.\footnote{
For a small delta-function impulse, a load resistance much smaller than the TES resistance at the working point, $R_\mathrm{L} \ll R_0$, and a  rise time much faster than the decay time, the latter approaches the effective time constant $\tau_\mathrm{decay} \to \tau_\mathrm{eff} \approx \tau (1 + \beta_I) / (1 + \beta_I +\mathcal{L}_I)$, where $\tau = C/G$ is the ratio between the heat capacity and the thermal conductance, and $\beta_I = \partial\ln R/\partial \ln I$ is the current sensitivity evaluated at the temperature $T_0$ at the WP~\cite{2005cpd..book...63I}.
The bias current for TES1 at the WP is a factor $\sim3$ larger than for TES2 (cf. Sec.~\ref{sec:tes}) leading to a higher Joule heating $P_{J_0}$ and, thus, a higher loop gain, which lowers the decay time, since the loop gain is given by $\mathcal{L}_I = P_{\mathrm{J}_0} \alpha_I / (G T_0)$  with $\alpha_I  = \partial\ln R/\partial \ln T$ the temperature sensitivity at the WP bias current.
}  
The distributions of the rise times are less symmetric for both modules and wavelengths and 
show a clear skew towards lower values, a behaviour observed previously~\cite{Schwemmbauer:2025evp}.
Possible reasons are that $\tau_\mathrm{rise}$ is constrained to be positive and is correlated with the pulse height and decay time.  
Furthermore, whereas the mean values coincide for TES1 for pulses at different wavelengths, a systematic shift towards larger rise and decay times can be observed for TES2 for pulses with shorter wavelength. 
This leads to an overall broadening of the joint distribution for TES2 and an overall worse description of the parameter distribution with a multivariate Gaussian. 
This can also be seen when comparing the contour lines of constant likelihood in both panels in the top row of Fig.~\ref{fig:photonness} and the histograms of the likelihood values defined in Eq.~\eqref{eq:lnl-gamma}, shown in the lower panels. 
For TES1 (TES2), the minimum likelihood value is around $-10$ ($-5$), which indicates an overall better agreement between model and data for TES1. 
For this reason, we decide to use a lower threshold $\xi$ for TES2: 
the broader distributions are expected to lead to larger background contamination which is reduced by lower values of $\xi$ 
even though this will also reject more true photon events.
In future work, we will explore going beyond multivariate normal distributions to describe the observed distribution of parameters.

The observed energy dependence of the rise and decay times 
suggests that optimal filtering~\cite[e.g.,][]{2016JLTP..184..374F}--essentially fitting an average pulse shape with varying amplitude--might not be well-suited for the energy range of our data set; 
one of its underlying assumptions is that the pulse shape is independent of deposited energy. 

With the chosen thresholds, we find on average an analysis efficiency of $\epsilon_a = 94.5\,\%$ (91.0\,\%), i.e., the fraction  of all traces in the calibration data set are identified as photon like for TES1 (TES2) in the energy range from 1.175 to 1.893\,eV.
We now use these photon-like events to perform the energy calibration of our modules and derive their energy resolution.

\subsection{Energy Calibration and Resolution}
\label{sec:energy_calibration}

\begin{figure*}[htb]
    \centering
    \includegraphics[width=0.49\linewidth]{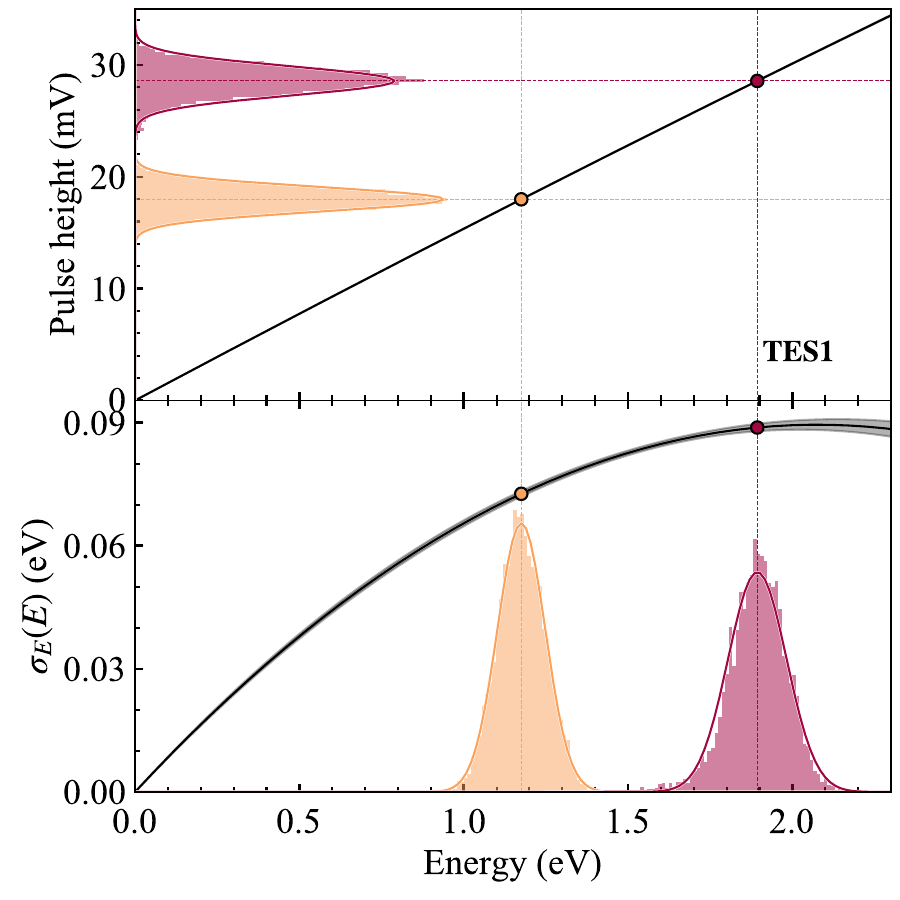}
    \includegraphics[width=0.49\linewidth]{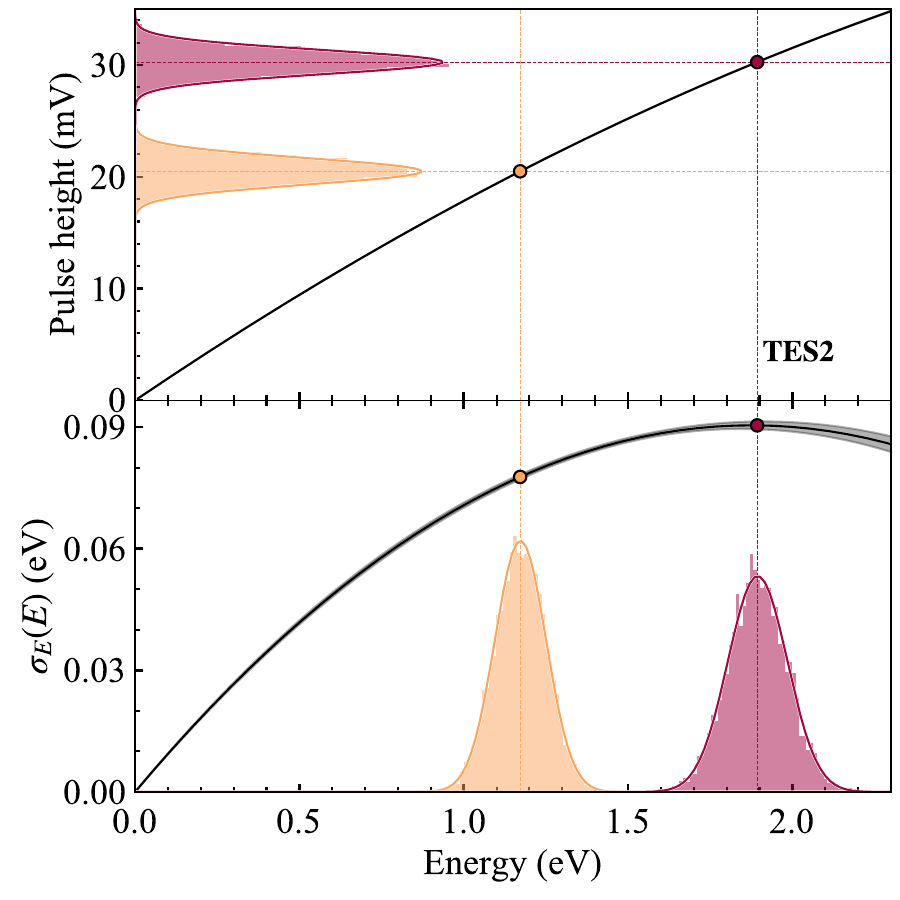}
    \caption{Energy calibration and resolution for both TES1 (left panels) and TES2 (right panels). \textit{Top:} Heights of the pulses of the calibration data sets as a function of laser energy. 
    Histograms along the $y$~axis show the observed pulse-height distributions. 
    The error bars on the mean values of the pulse heights are too small to be seen. The mean values are fit with a second degree polynomial with intercept zero which we use for our energy calibration.
    \textit{Bottom:} Standard deviations of the energy distributions as a function of laser energy. The functional dependence is again described with a second degree polynomial with intercept zero. 
    Histograms along the $x$~axis show the observed energy distribution derived from the pulse heights. 
    }
    \label{fig:energy-cal-res}
\end{figure*}

We perform the energy calibration of our sensors with the photon-like events in the calibration data set (cf. Tab.~\ref{tab:cal-dataset}).
These events should correspond to an energy deposition of $\sim1.175$~eV and $\sim1.893$~eV, respectively, depending on the laser used. 
We use the height of the fitted pulses as our energy estimator 
rather than the pulse integral, as this choice results in an improved energy resolution~\cite{RubieraGimeno:2025bhr}
(note that we do not fix the pulse shape).
For each trace, we compute the pulse height from the best-fit parameters of our fitting function in Eq.~\eqref{eq:pheno-func}, 
\begin{equation}
    h = -2V_0 \frac{\tau_\mathrm{rise}}{\tau_\mathrm{rise} + \tau_\mathrm{decay}}\left(\frac{\tau_\mathrm{decay}}{\tau_\mathrm{rise}}\right)^{\frac{\tau_\mathrm{decay}}{\tau_\mathrm{rise} + \tau_\mathrm{decay}}},
\end{equation}
and show the distribution of pulse heights in the upper panels of Fig.~\ref{fig:energy-cal-res}.
The two distributions are well described with normal distributions.
Similar to Ref.~\cite{Schwemmbauer:2025evp}, we fit the mean values of the two distributions as a function of energy.
We opt for a second degree polynomial with intercept equal to zero (black lines in the top panels of Fig.~\ref{fig:energy-cal-res}), in order to account for nonlinearities in the TES response that might occur for high energy depositions. 
As we only have two available energies, the parameters of the polynomial are fully determined. 
By inverting the polynomial, we assign an energy to each fitted pulse. 

The resulting distributions of energy are shown in the lower panels of Fig.~\ref{fig:energy-cal-res}, which are again well described by normal distributions with mean $\mu_E$ and standard deviations $\sigma_E$. 
From these distributions, we can derive the energy resolution of our detectors, which we report in Tab.~\ref{tab:energy-resol} both in terms of the ratio $\sigma_E / \mu_E$ and the full width half maximum (FWHM).
Both modules have comparable energy resolution of 6.2\,\% and 6.6\,\% at 1.175\,eV  that slightly improves to roughly 5\,\% at 1.893\,eV.
Using the integral of the pulses instead of their heights degrades the energy resolution at 1.175\,eV to 9.2\,\% (12.3\,\%) for TES1 (TES2).  
These numbers are compatible with previous results~\cite{RubieraGimeno:2023} and 
highlight again that the pulse height is the favorable choice to estimate the deposited photon energy.

\begin{table}[htb]
    \centering
    \caption{Energy resolution of our TES modules at different laser energies.}
    \begin{tabular}{c|c|c}
    \hline
    \hline
    Energy (eV) & $\sigma_E / \mu_E$ & FWHM (eV) \\
    \hline
    \multicolumn{3}{c}{TES1} \\
    \hline
    1.175 & $0.0618 \pm 0.0006$ & $0.172 \pm 0.002$ \\
    1.893 & $0.0469 \pm 0.0005$ & $0.210 \pm 0.002$ \\
    \hline
    \multicolumn{3}{c}{TES2} \\
    \hline
    1.175 & $0.0660 \pm 0.0005$ & $0.183 \pm 0.001$ \\
    1.893 & $0.0474 \pm 0.0005$ & $0.211 \pm 0.002$ \\
    \hline
    \end{tabular}
    \label{tab:energy-resol}
\end{table}

To find the energy resolution at different energies, $\sigma_E(E)$, we assume that it also follows a second order polynomial with intercept zero.
The result is shown as a black line with grey uncertainty band in the lower panels of Fig.~\ref{fig:energy-cal-res}. 
The functional dependence $\sigma_E(E)$ is used in the next section to derive the full energy dispersion matrix of our sensors. 
We have checked that using instead a first degree polynomial with non-zero intercept does not change our results for the measured background spectrum presented in Sec.~\ref{sec:bkg}.

\subsection{Detector Response Matrix}
\label{sec:drm}

\begin{figure}
    \centering
    \includegraphics[width=0.9\linewidth]{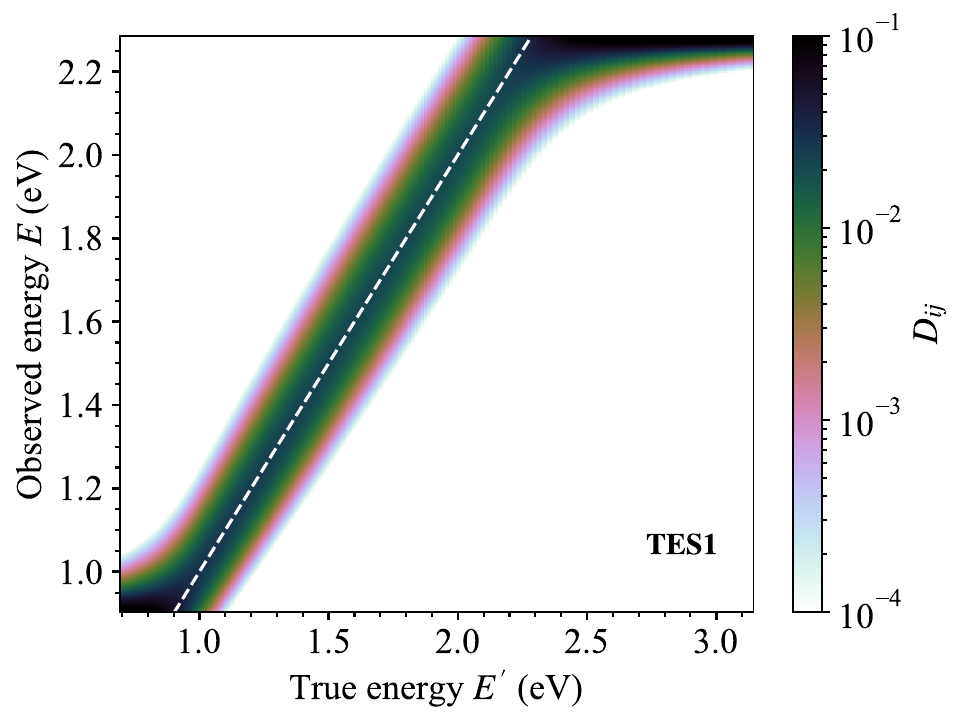}
    \caption{Energy dispersion matrix for TES1 derived from the calibration data. The white dashed line indicates where the observed energy is equal to the true energy, $E=E'$.}
    \label{fig:edisp}
\end{figure}

The observed photon flux at the TES is the result of folding the true photon flux with the detector response matrix that includes the system detection efficiency $\eta$ as well as the energy dispersion. 
The energy dispersion, $D(E, E')$, gives the probability that a photon of true energy $E'$ is observed at energy $E$. 
Motivated from the results in Sec.~\ref{sec:energy_calibration}, we assume that it follows a Gaussian distribution. 
Using the energy resolution curve $\sigma_E(E)$ from Fig.~\ref{fig:energy-cal-res}, 
we construct the energy dispersion matrix, which is normalized over the observed energy axis (index $i$),
\begin{equation}
    D_{ij} \equiv D(E_i, E'_j) = \frac{\exp\left[-\frac{1}{2}\left(\frac{E_i-E'_j}{\sigma_E(E_i)}\right)^2\right]}{\sum_i \exp\left[-\frac{1}{2}\left(\frac{E_i-E'_j}{\sigma_E(E_i)}\right)^2\right]}.
    \label{eq:edisp}
\end{equation}
The energy dispersion matrix for TES1 for the observed energy ranges of the calibration data is shown in Fig.~\ref{fig:edisp}. For most observed energies, $D_{ij}$ peaks along the line where $E=E'$. This is not true at the edges of the observed energy range, where the normalization (the denominator) in Eq.~\eqref{eq:edisp} broadens the energy dispersion significantly.

For the full detector response, we also need to account for the SDE as a function of photon energy.
For a single energy $E_0 = 1.17\,$eV, the SDE $\eta_0$ has been measured as described in Sec.~\ref{sec:sde}.
In general, the SDE will depend on photon energy due to (a) the energy-dependent probability, $\eta_\mathrm{abs}(E)$, that a photon is absorbed by the TES; (b) the energy-dependent transmission inside the optical fiber, $\eta_\mathrm{fiber}(E)$; and (c) energy-dependent transmission due to curling of the fiber, $\eta_\mathrm{curling}(E)$.
Following Ref.~\cite{RubieraGimeno:2025bhr}, we take $\eta_\mathrm{abs}(E)$ and $\eta_\mathrm{fiber}(E)$ from the literature~\cite{Dreyling-Eschweiler:2014eya,2019LaPhL..16c5108A}.
Bends or curls in the fiber will lead to radiation loss particularly at longer wavelengths as the condition for total internal reflection inside the fiber is no longer satisfied.
Curling of the fiber inside the cryostat is unavoidable and the exact curvature radius and number of loops is difficult to measure precisely. 
Here, we model the resulting efficiency from curling with the help of the logistic function of the form
\begin{equation}
    \eta_\mathrm{curling}(E) = \left(1 - \frac{1}{1 + \exp\left(\frac{E - E_\mathrm{min}}{a}\right)}\right)^\ell,
\end{equation}
with $E_\mathrm{min}$ the energy where the logistic function is equal to $1/2$.
Both $E_\mathrm{min}$ and $a$ are related to the radius of curvature of the fiber curling loops whereas $\ell$ represents the number of loops~\cite{RubieraGimeno:2025bhr}. 
In practice, we determine the curling parameters from the measured spectrum, see Sec.~\ref{sec:llh}.

\begin{figure}
    \centering
    \includegraphics[width=0.95\linewidth]{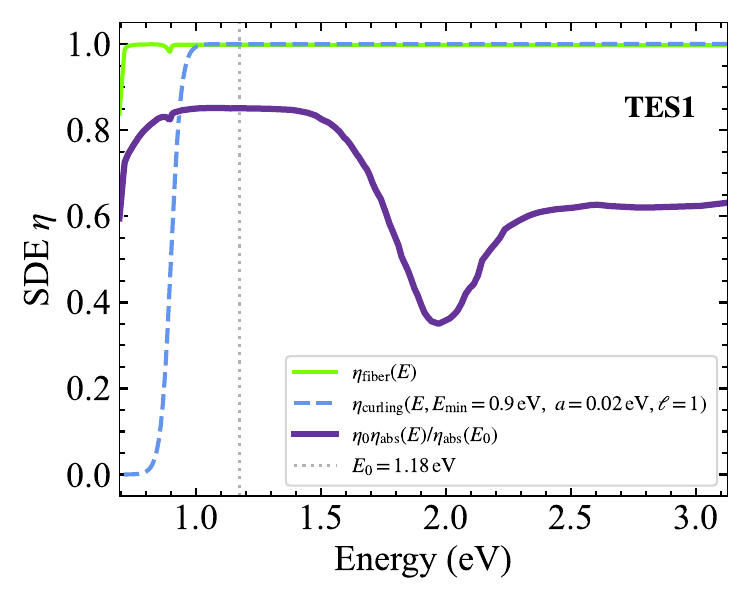}
    \caption{Energy dependent system detection efficiency for TES1. 
    The contributions from fiber transmission and fiber curling losses as well as the absorption probability of a photon in the optical stack including the TES are taken from the literature. 
    The total SDE (dark purple line) is scaled to the observed value measured in Sec.~\ref{sec:sde}. 
    }
    \label{fig:sde}
\end{figure}

The full energy-dependent detection efficiency is then 
\begin{equation}
 \eta(E) = \eta_0 \frac{\eta_\mathrm{abs}(E) \eta_\mathrm{fiber}(E) \eta_\mathrm{curling}(E)}{\eta_\mathrm{abs}(E_0) \eta_\mathrm{fiber}(E_0) \eta_\mathrm{curling}(E_0)},
 \label{eq:sde-energy-dep}
\end{equation}
where the denominator ensures that $\eta(E)$ is equal to the measured value $\eta_0$ at energy $E_0$.
We show the energy-dependent SDE and its individual components for example values of $E_\mathrm{min}, a$, and $\ell$ in Fig.~\ref{fig:sde}.

Lastly, the full detector response $R(E',E)$ (in units of area and solid angle) is found by multiplying the efficiency from Eq.~\eqref{eq:sde-energy-dep} with the energy dispersion matrix [Eq.~\eqref{eq:edisp}] and the area and solid angle of our single-mode optical fiber, 
\begin{equation}
    R(E, E') = D(E, E')\eta(E') (\pi r_\mathrm{core}\,\mathrm{NA})^2,
    \label{eq:irf}
\end{equation}
where $r_\mathrm{core} = 3.1\,$µm is the fiber core radius, and $\mathrm{NA} = 0.14$ the numerical aperture of the fiber~\cite{RubieraGimeno:2025bhr}.

\section{The background dark count rate}
\label{sec:bkg}

For both detectors, we have measured the extrinsic background dark count rate (bDCR) where the optical fiber coming from the attenuation stage at MS\,B (cf. Fig.~\ref{fig:eff-setup}) is disconnected and MS\,B is covered with a metal cap. 

\begin{figure*}[htb]
\centering
\includegraphics[width=.48\linewidth]{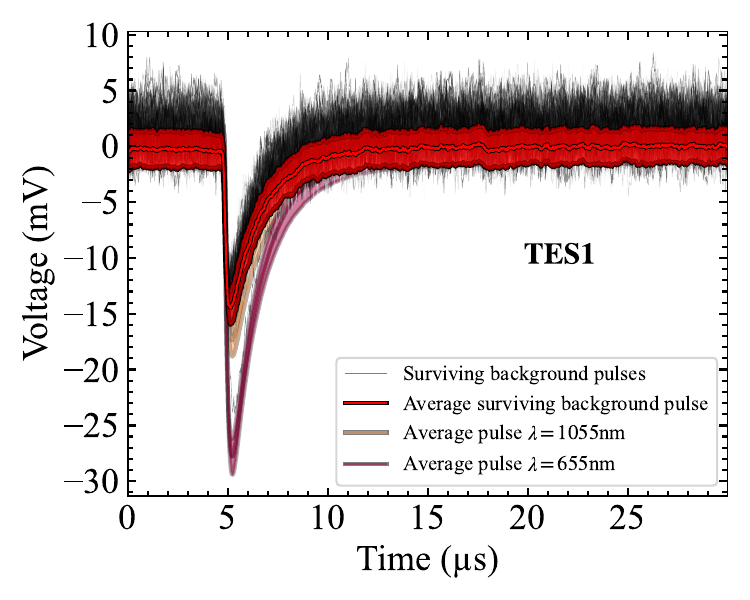}
\includegraphics[width=.48\linewidth]{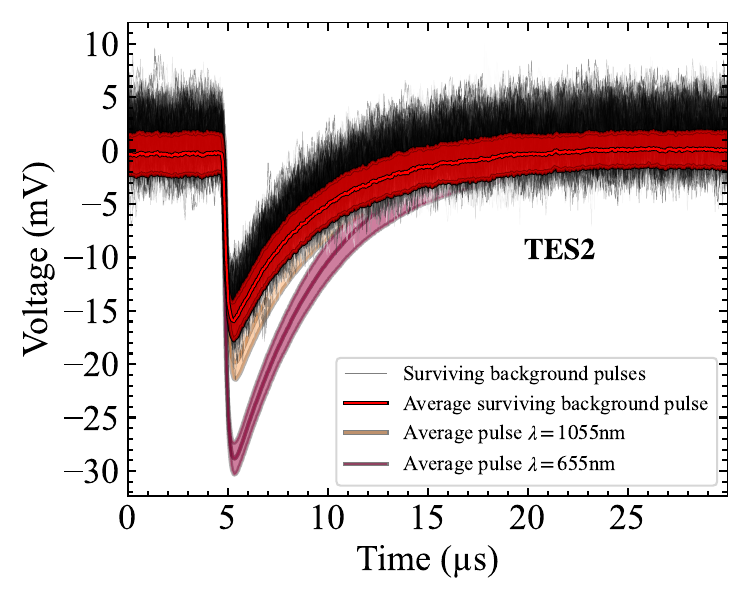}
\caption{Background pulses surviving the photon selection for TES1 (left) and TES2 (right).
Average pulses for the surviving background as well as for the calibration data sets are shown as well. 
Comparing the height of the average background pulse with the height of average pulse from the 1058\,nm laser, it is clear that most of the background pulses deposit energies $<1.17\,$eV. 
}
\label{fig:surv-pulses}
\end{figure*}

For TES1 (TES2), 19 (20)\, hours of data were recorded. 
Just as the calibration data, these data were taken in triggered mode with the trigger threshold set to 12\,mV.
Each pulse is fitted with the phenomenological function $V(t)$ of Eq.~\eqref{eq:pheno-func}. 
Only pulses with fit parameters resulting in likelihoods $\ln\pazocal{L}_\gamma < \xi$, which indicates photon-like events, are retained.
The full distributions of $\ln\pazocal{L}_\gamma$ values are shown in the lower panels of Fig.~\ref{fig:photonness} and the traces of the 
surviving pulses are presented in Fig.~\ref{fig:surv-pulses}. 
Clearly, the surviving pulses appear photon-like and the deposited energies are mostly below 1.17\,eV. 

The number and rate of triggers before and after photon selection are reported in Tab.~\ref{tab:bkg}.
For TES2, the background rate is found to be a factor of $\sim2.2$ higher than for TES1. 
The most likely reason are the broader distribution of pulse parameters for TES2 discussed in Sec.~\ref{sec:photonness} leading to a larger contamination of background events. 
An additional contribution to the higher background rate could arise from the trigger threshold, which is the same for both sensors. 
From the energy calibration curves in Fig.~\ref{fig:energy-cal-res} these trigger thresholds correspond to 0.78\,eV (0.65\,eV) for TES1 (TES2). 
However, below $\sim$0.9\,eV, the photon flux is anyway heavily suppressed predominantly due to fiber curling (see Fig.~\ref{fig:sde}).
Another reason could be differences in the fiber curling or a change in temperature between the two measurements which were taken several days apart.
As we show in the next section, by fitting the recorded photon spectra using an unbinned likelihood framework, the temperatures are compatible with each other. However, the reconstructed fiber curling parameters appear different.

\begin{table}[htb]
    \centering
    \caption{Characteristics of bDCR measurements before and after photon selection. Uncertainties are derived from Poissonian statistics.
    }
    \begin{tabular}{c|cc}
\hline
\hline
    & TES1 & TES2 \\
\hline
$T_\mathrm{obs}$ (hr)   & 19 &  20  \\
Number of recorded pulses & $1524 \pm 39$ & $2114\pm46$\\
Raw trigger rate ($10^{-2}$ s$^{-1}$) & $2.23 \pm 0.06$ & $2.94\pm0.06$\\
Surviving pulses   & $167\pm12.9$ &  $363 \pm 19.1$  \\
bDCR ($10^{-3}$ s$^{-1}$) & $2.44 \pm 0.18$  & $5.04 \pm 0.26$ \\
\hline
    \end{tabular}
    \label{tab:bkg}
\end{table}

\subsection{Unbinned likelihood framework}
\label{sec:llh}
We introduce an unbinned likelihood framework, which is based on the extended maximum likelihood method~\cite{Barlow:1990vc}, to determine the spectrum of the bDCR.
After calibration, each TES pulse has an assigned \textit{observed} energy $E_i$, $i=1,\ldots,N$, and time $t_i$, which we take as the start of trigger window. 

Given a photon source characterized by its spectral intensity $\phi(E', t)$, measured in photons per unit time, energy interval, unit area, and solid angle, which depends on true photon energy $E'$, time $t$, and model parameters $\boldsymbol{\theta}$, we determine the expected photon rate per observed energy and time interval, $dn/dE$, at observed energies $E_i$ and time $t_i$. This is achieved by folding $\phi(E', t)$ with the detector response matrix $R$ from Eq.~\eqref{eq:irf},
\begin{equation}
    \frac{dn}{dE}(E_i, t_i) = \sum_j R(E_i, E'_j) \phi(E'_j, t_i).\label{eq:folding}
\end{equation}
The total number of expected counts, $N_\mathrm{exp}$, is obtained by integrating $dn/dE$ over the observed energy and time ranges. 

In the limit of infinitesimally small bins in observed energy and time, where each bin contains either zero or one event, the unbinned form of the Poisson likelihood can be expressed as~\cite{Barlow:1990vc},
\begin{equation}
\ln\pazocal{L}(\boldsymbol{\theta}|\{E_i, t_i\}) = -N_\mathrm{exp}(\boldsymbol{\theta}) + \sum_i \ln \left(\frac{dn}{dE}(E_i, t_i; \boldsymbol{\theta})\right).
\label{eq:llh}
\end{equation}
The unbinned likelihood formulation ensures maximum use of information and no dependency of the results on a chosen binning. 

The major source of background is expected to originate from high-energy tail of blackbody radiation (BBR) from the lab environment at $\sim300\,$K that inevitably couples into the optical fiber~\cite{Rosenberg:2005zme, RubieraGimeno:2025bhr,Rivasto:2025gjk}. 
Therefore, we fit the observations with the spectral photon intensity of blackbody radiation at temperature $T$, 
\begin{equation}
    B(E',T) = \frac{2}{h^3 c^2}\frac{E'^2}{\exp\left(-\frac{E'}{k_BT}\right) - 1},\label{eq:bbr}
\end{equation}
where $k_B$ is the Boltzmann constant.
We assume the temperature, and therefore $B(E',T)$, to be constant in time.
Thus, our model has in total 4 parameters: the temperature and the fiber curling parameters, so that $\boldsymbol{\theta} = (T, E_\mathrm{min}, a, \ell)$. 
The loop with the minimum curling radius will determine the minimum energy cutoff $E_\mathrm{min}$. 
Within the cryostat, we roughly have a quarter turn ($\ell \approx 1/4$) with curling radius of $\sim2$\,cm when the fiber is plugged into the TES. 
From previous work, we expect that this curling radius results in  $E_\mathrm{min} \sim 0.9\,$eV~\cite{RubieraGimeno:2025bhr}.
In practice, we fix $\ell=1/4$, as this parameter is highly degenerate with the other model parameters.
The remaining parameters are found by maximizing the likelihood in Eq.~\eqref{eq:llh} numerically using \texttt{Minuit}.

\subsection{Results}
\label{sec:bkg-result}

\begin{figure*}[htb]
    \centering
    \includegraphics[width=0.49\linewidth]{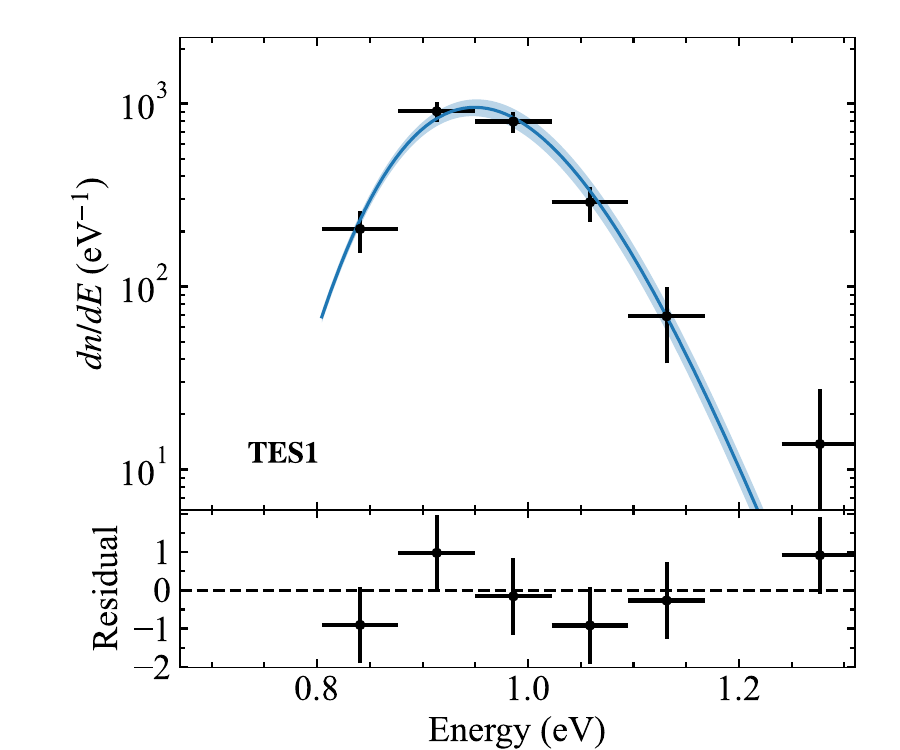}
    \includegraphics[width=0.49\linewidth]{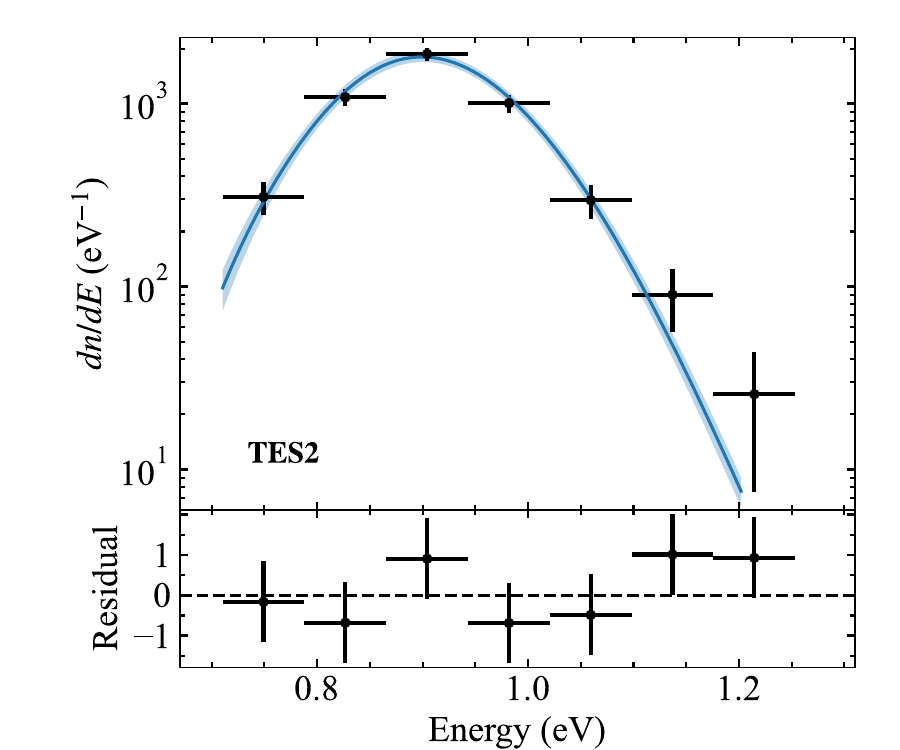} \\
    \includegraphics[width=0.49\linewidth]{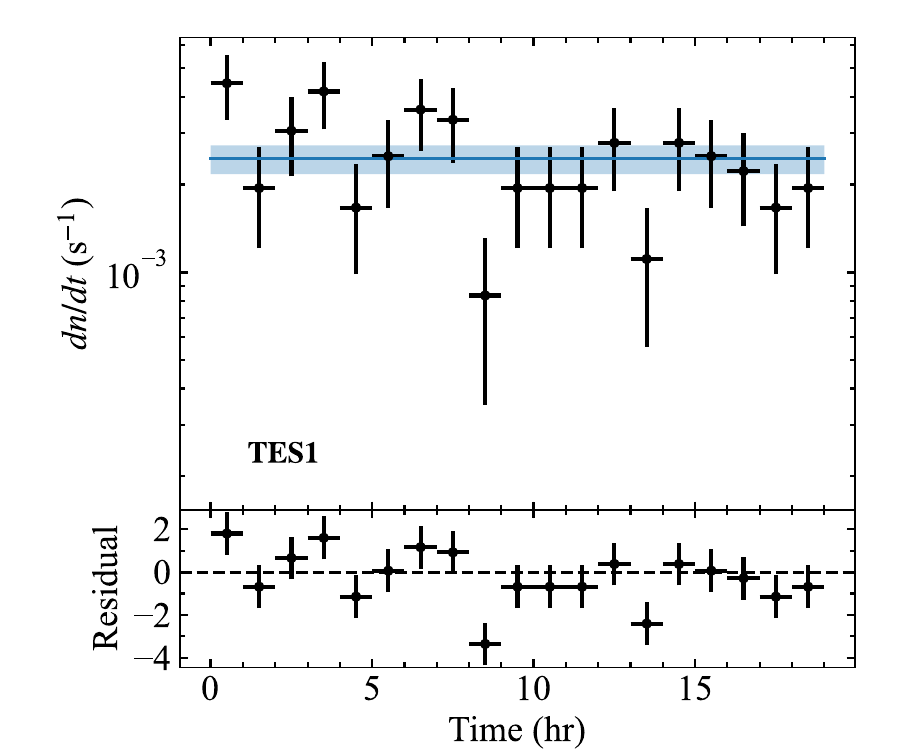}
    \includegraphics[width=0.49\linewidth]{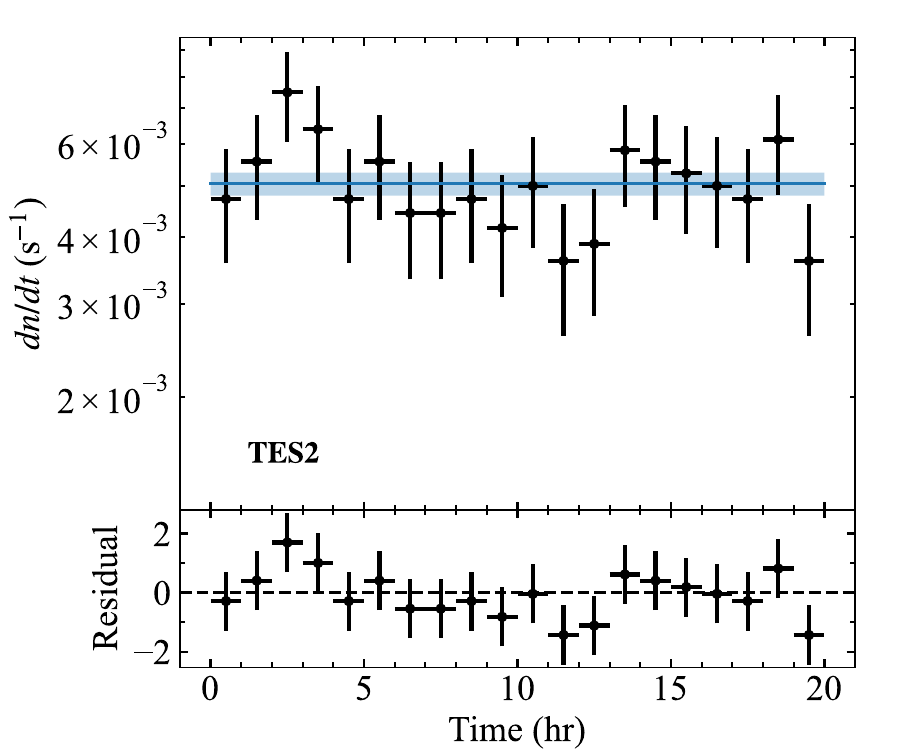} \\
    \caption{\textit{Top:} Energy spectra and best-fit spectral models for TES1 (left) and TES2 (right). The chosen binning of the individual pulses corresponds to the energy resolution $\sigma_E$ at 1.17\,eV.
    \textit{Bottom:} Reconstructed count rates and best-fit constant model count rate. A binning equal to one hour is chosen.}
    \label{fig:spec-fit}
\end{figure*}

Figure~\ref{fig:spec-fit} shows the observed photon energy spectra (top row) and the count rate as a function of time (bottom row) for both TESs along with the best-fit models. 
The best-fit parameters are reported in Tab.~\ref{tab:fit}, together with $\chi^2$ values. 
We note that the $\chi^2$ values for both the energy and time domain are computed after the fit by binning the data. 
For the energy spectra, a bin width equal to the energy resolution at 1.17\,eV is chosen. For the count rates as a function of time, the bin width is set to one hour. 

In agreement with Ref.~\cite{RubieraGimeno:2025bhr}, both spectra are well described by a BBR background with the fiber-curling low-energy cutoff.
The reconstructed temperatures of
$297.4^{+2.0}_{-1.8}$\,K and $295.4^{+1.6}_{-1.9}$\,K
for both modules are compatible with each other within uncertainties and are in agreement with our lab environment. 
The main difference between the two modules arises for energies below 1\,eV, where the model is dominated by the cutoff. 
The cutoff for TES1 is fitted to be steeper (corresponding to smaller values for $a_\mathrm{exp}$) at slightly higher cutoff energies. 
For TES1, estimating the exact uncertainty of the $a_\mathrm{exp}$ parameter is difficult. This is likely due to the overall lower fit quality (discussed below) and the effectively larger energy threshold for TES1 discussed in Sec.~\ref{sec:bkg} above, leading to fewer counts at lower energies. 
Therefore, the low uncertainties in Tab.~\ref{tab:bkg} should be regarded with caution.
In the future, we plan to determine the fiber transmission at long wavelength independent of the TES, which would then allow us to fix these parameters during the fit. 
In this way, one could also determine whether the lower count rates below 1\,eV are due to fiber curling or a lower sensitivity of this TES module at longer wavelengths.

From the residuals in Fig.~\ref{fig:spec-fit} and the derived $\chi^2$ values, it is also obvious that the fit quality is lower for TES1. 
The lower fit quality in the energy domain is mainly driven by two high energy photon-like events, reconstructed at 1.3\,eV and 1.7\,eV, respectively (the 1.7\,eV photon is also clearly visible in Fig.~\ref{fig:surv-pulses} as a high amplitude pulse). 
In the time domain, one can observe a significant drop in the count rate around 8\,hours leading to a comparatively large $\chi^2_t$ value. 

\begin{table}[htb]
    \centering
    \begin{tabular}{c|cc}
    \hline
    \hline
    Parameter & TES1 & TES2 \\
    \hline
$T$ (K)               & $    297.37^{+1.96}_{-1.79}$ & $    295.40^{+1.58}_{-1.94}$ \\
$E_\mathrm{min}$ (eV) & $     0.943^{+0.006}_{-0.003}$ & $     0.915^{+0.007}_{-0.007}$ \\
$a_\mathrm{exp}$ ($10^{-3}$ eV) & $      2.82^{+0.06}_{-0.06}$ & $      4.44^{+0.25}_{-0.25}$ \\
$\chi^2_E$   / d.o.f. & $      4.55~/~ 4$ & $      3.88~/~ 4$ \\
$\chi^2_t$   / d.o.f. & $     30.80~/~16$ & $     12.41~/~17$ \\

    \hline
    \end{tabular}
    \caption{Best-fit parameters for the BBR background model derived from our unbinned likelihood framework. Also reported are the $\chi^2$ values and the numbers of degrees of freedom (d.o.f.) for both the energy ($\chi^2_E$) and time domain ($\chi^2_t$). These are determined after the fit by binning the data as shown in Fig.~\ref{fig:spec-fit}.}
    \label{tab:fit}
\end{table}

In Ref.~\cite{RubieraGimeno:2025bhr} an excess was observed above the BBR background around $\sim1.16\,$eV.
Even though our TESs are different from the one used in Ref.~\cite{RubieraGimeno:2025bhr}, we test whether we observe a similar excess. 
To this end, in addition to the BBR, we add a monochromatic spectral line to the model, 
\begin{equation}
    \phi_\mathrm{line}(E') = n_0\delta(E'-E'_\mathrm{line}), \label{eq:spectral-line}
\end{equation}
where $n_0$ is the flux of signal photons in units of photons per second (note that the delta function is in units per energy), 
that we assume to be constant in time. 
In practice, we use a discretized version\footnote{
The discretized version is given by
\begin{equation}
    \phi_\mathrm{line}(E') = \frac{n_0}{\Delta E'} \delta_{jk},
\end{equation}
where the Kronecker delta is equal to 1 for the true energy $E'_k$ that is within the true energy bin containing  $E'_\mathrm{line}$, where we used finely binned intervals $\Delta E'$.}
and fold it with our instrumental response function, see Eq.~\eqref{eq:folding}. 
For a fixed line energy, we maximize the likelihood of Eq.~\eqref{eq:llh} in terms of the BBR temperature, the fiber-curling parameters, and the flux normalization of the line, $n_0$, and compute the log-likelihood difference, referred to as test statistic TS, for the two tested hypotheses $\mathrm{TS} = -2(\ln\pazocal{L}_0 - \ln\pazocal{L}_1)$, where index 0 (1) refers to the BBR-only (BBR and spectral line) model.
The model with the spectral line has one extra degree of freedom; thus the likelihood should asymptotically converge to a $\chi^2$ distribution with one degree of freedom~\cite{Wilks:1938dza}. 
For TES1 (TES2), we find a log-likelihood difference of 1.54 (6.29) corresponding to a significance of the line component of 
1.2\,$\sigma$ (2.5\,$\sigma$).
The extra component for TES2 has a best-fit value of $n_0 = (1.72 \pm 0.93)\times 10^{-4}$\,Hz.

\section{Sensitivity for rare-event signals at 1064nm}
\label{sec:sensitivity}

With the determined background, we can now estimate what fluxes of rare events our TES could potentially measure. 
We focus on applications in an LSW experiment or an axion interferometer introduced in Sec.~\ref{sec:intro}.
Both experiments share the common feature that the signal is expected to be a low flux of monochromatic photons at a wavelength of roughly 1064\,nm. 

We therefore proceed to simulate such a signal on top of a BBR background.
Motivated by the ALPS~II design goals, we assume an observation time $t_\mathrm{obs} = 20\,$days and use the detector response found for TES1 for simplicity.
On top of the BBR, we add a spectral line as in Eq.~\eqref{eq:spectral-line} to the model that emulates our signal.
No additional spectral-line background contribution is included here.  
We test 7 values of $n_0$ between $1.4 \times 10^{-5}\,$Hz and $9.2 \times 10^{-5}\,$Hz, motivated by the 
ALPS~II design sensitivity of $3\times 10^{-5}\,$Hz~\cite{2013JInst...8.9001B,2025arXiv251214110B} (see also below). 

\begin{figure}
    \centering
    \includegraphics[width=0.9\linewidth]
    {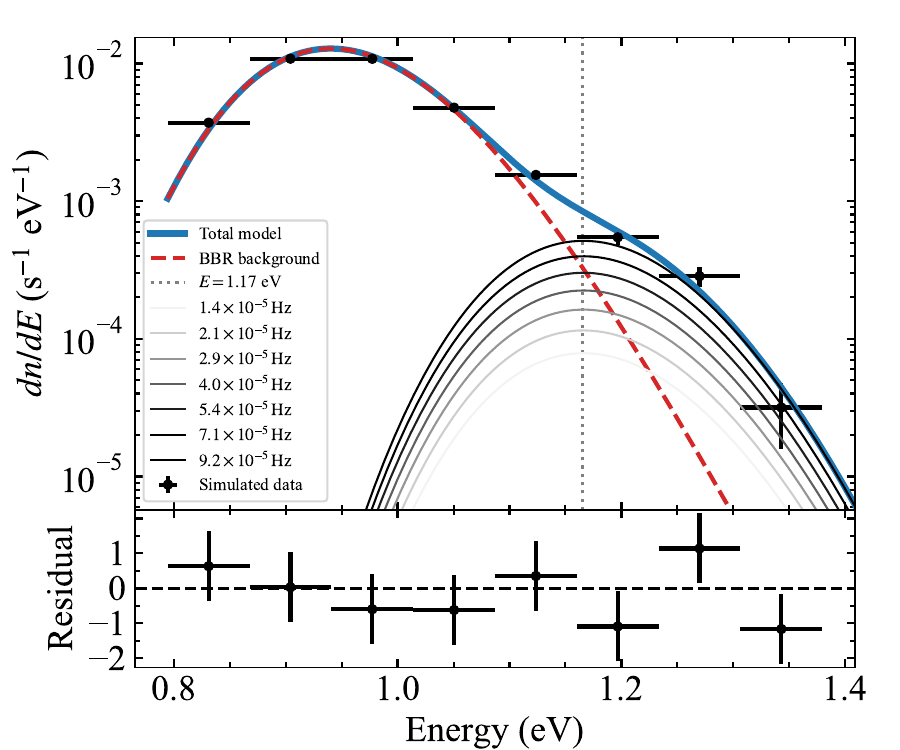}
    \caption{
    Simulated data set with injected signal line at 1.17\,eV convolved with the IRF. The model line is shown for different rates $n_0$.
    The total signal model is the sum of the line and BBR contributions. 
    The maximum value of $n_0$ tested here is assumed for the simulation, the other tested values of $n_0$ are shown as gray lines for visualization. 
    }
    \label{fig:sim-data}
\end{figure}

To evaluate which values of $n_0$ we could significantly detect in a measurement, we generate Poisson distributed Monte Carlo data sets. 
To this end, we inject signals for different values of $n_0$ on-top of a BBR background with best-fit temperature $T$ and fiber curling parameters as determined for TES1 in Sec.~\ref{sec:bkg-result}.
We draw $300$ data sets from a Poisson distribution centered around the expected number of photons, $N_\mathrm{exp}$, for each injected signal flux $n_0$. 
For each event in each data set, we draw an arrival time from a uniform distribution between $[0, t_\mathrm{obs})$ and an energy from a distribution generated from the assumed model spectrum. 
An example of such a data set is shown in Fig.~\ref{fig:sim-data} for the highest value of $n_0$ assumed here. 
The injected spectral line leads to a clear excess at energies $\gtrsim 1.1$\,eV over the BBR background. 
In Fig.~\ref{fig:sim-data}, we also show spectral lines for the other injected line strengths. 
Note that the error bars for the data points are much smaller in Fig.~\ref{fig:sim-data} compared to Fig.~\ref{fig:spec-fit} as we have assumed an observation time of 20 days, i.e., $\sim$24 times longer than the background data discussed in Sec.~\ref{sec:bkg}.
It has already been demonstrated that our detectors can be operated for such prolonged periods of time~\cite[see, e.g.,][]{2022JLTP..tmp...93S, Schwemmbauer:2025evp}. 
Likewise, the first science campaign of ALPS~II consisted of two extended data taking runs over a 
total period of $\sim 4\,$months. 

For each injected value of $n_0$ and pseudo experiment, we perform two fits with our unbinned likelihood framework. 
The fitted model now consists of the BBR background and a spectral line with fixed line energy at 1.17\,eV (the line energy is known in the experiment) and free normalization, which we refer to as $n_0^\mathrm{fit}$ in order to distinguish it from the injected value $n_0$. 
We proceed as in Sec.~\ref{sec:bkg-result} and fit a BBR model only by fixing $n_0^\mathrm{fit} = 0$, and find the maximum likelihood $\ln\pazocal{L}_0$.
Next, we free $n_0^\mathrm{fit}$ and find the unconditionally maximized likelihood, $\ln\pazocal{L}_1$, and compute again test statistic $\mathrm{TS} = -2(\ln\pazocal{L}_0 - \ln\pazocal{L}_1)$ and the significance for the additional line component.  
A $\mathrm{TS}$ value $\gtrsim 25$ indicates that we have detected the line at a significance of $5\,\sigma$.

\begin{figure}
    \centering
    \includegraphics[width=0.9\linewidth]{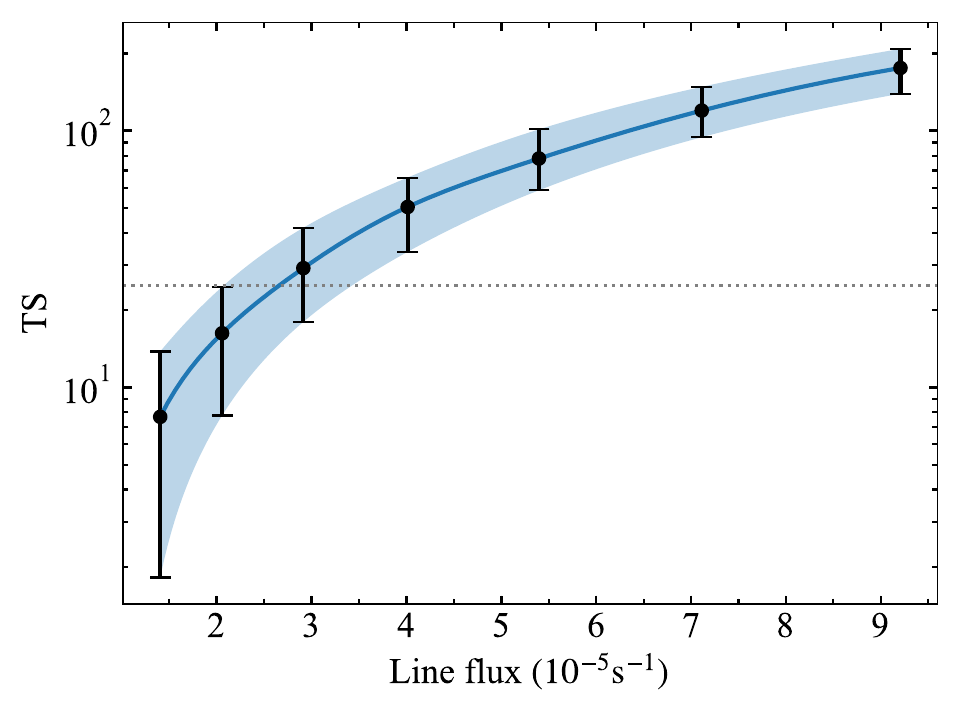}
    \caption{Median TS values from comparing the fit without a line with the fit including a line signal for 300 pseudo experiments for each injected line flux $n_0$. Error bars are found from the 16\,\% and 84\,\% quantiles of the TS distribution. The blue line and shaded region indicates a cubic spline interpolation of the median and quantile values, respectively. }
    \label{fig:ts-vs-flux}
\end{figure}

For each injected line flux, we show the median of the $\mathrm{TS}$ value distribution of the 300 pseudo experiments together with the 16\,\% and 84\,\% quantiles in Fig.~\ref{fig:ts-vs-flux}.
From a cubic spline interpolation to the TS vs. flux dependence, we find that the flux that can be detected at $ \geqslant 5\sigma$ significance within a 20\,day measurement is given by
\begin{equation}
     n_0 \geqslant 2.66_{-0.59}^{+0.77} \times10^{-5}\,\mathrm{Hz}, 
     \label{eq:flux_thr}
\end{equation}
which corresponds to an optical power of $(4.97_{-1.09}^{+1.43})\times10^{-24}\,\mathrm{W}$ at 1064\,nm.
This result might come as a surprise: even though the background rate is above $10^{-3}\,$Hz we can still detect a signal at 1.17\,eV with a flux as low as $\sim 3\times10^{-5}$\,Hz. 
The reason is the energy resolution capability of our detector that is fully exploited in our unbinned likelihood framework. 

Finally, we determine what value of the photon-axion coupling this flux corresponds to. 
As examples, we consider the ALPS~II experiment at design sensitivity as well as a LIDA-like axion interferometer~\cite{2024PhRvL.132s1002H}. 
In an LSW experiment, in the limit of small WISP masses, $m_a\lesssim 10^{-5}\,$eV, and neglecting gaps between dipole magnets, the photon-WISP mixing probability (in SI units) in one cavity is given by 
\begin{equation}
    P_{a\leftrightarrow\gamma} = \left(\sqrt{\frac{\hbar c}{\mu_0}}\frac{g_{a\gamma} B L N}{2}\right)^2
        ,
        \label{eq:pag}
\end{equation}
where $\hbar$ is the reduced Plank constant, $\mu_0$ is the vacuum permeability, $B$ is magnetic field strength, $L$ is length of one dipole magnet, and $N$ is the number of magnets.\footnote{
In suitable units, the prefactor $\sqrt{\hbar c / \mu_0} \approx 0.98999524 \,\mathrm{GeV}\,\mathrm{T}^{-1}\,\mathrm{m}^{-1} \approx 1\,\mathrm{GeV}\,\mathrm{T}^{-1}\,\mathrm{m}^{-1}$, which is a handy value to remember.
}
For production and regeneration cavities (PC and RC) with conversion probabilities $P^\mathrm{RC}_{a\leftrightarrow\gamma}$ and $P^\mathrm{PC}_{a\leftrightarrow\gamma}$ as well as power build-up factors $\beta_\mathrm{PC}$ and $\beta_\mathrm{RC}$, respectively, 
the expected signal photon rate is 
\begin{equation}
    n_\mathrm{LSW} = P^\mathrm{PC}_{a\leftrightarrow\gamma}
    P^\mathrm{RC}_{a\leftrightarrow\gamma} |\eta_P|^2\beta_\mathrm{PC}\beta_\mathrm{RC}\epsilon_d\epsilon_a\frac{P_\mathrm{laser}}{h\nu},\label{eq:lsw}
\end{equation}
where $|\eta_P|^2$ describes the spatial and spectral overlap between the WISP and the electric field mode~\cite{2010PhRvD..82k5018A}, and $P_\mathrm{laser}$ is the power of the laser operated at frequency $\nu$.
We take the values $N$, $B$, and $L$ from the ALPS~II RC~\cite{2025OExpr..3311153K}, and assume the ALPS~II design goals~\cite{2013JInst...8.9001B,2025arXiv251214110B}, i.e.,  a circulating power in the PC of $\beta_{PC}P_\mathrm{laser} = 150\,$kW, $|\eta_P|^2 = 0.9$, $\beta_\mathrm{RC} = 40{,}000$, and an overall detection efficiency of $\epsilon_d = 0.5$.
We also include the factor $\epsilon_a = 0.95$ which denotes the efficiency to retain photons in the analysis (see Tab.~\ref{tab:cal-dataset}).
This results in $n_\mathrm{LSW} = 1.57\times10^{-5}\,$Hz for $g_{a\gamma} = 2\times 10^{-11}\,\mathrm{GeV}^{-1}$. 
Plugging in $n_0$ from Eq.~\eqref{eq:flux_thr} for $n_\mathrm{LSW}$ in Eq.~\eqref{eq:lsw} and solving for $g_{a\gamma}$ with the help the definition of $P_{a\leftrightarrow\gamma}$ in Eq.~\eqref{eq:pag}, we find the photon-WISP coupling we could still significantly detect in ALPS~II to be
\begin{equation}
    g_{a\gamma}^\mathrm{ALPS~II} \gtrsim 
    2.38^{+0.16}_{-0.14}\times10^{-11}\,\mathrm{GeV}^{-1},
\end{equation}
which is close to the ALPS~II design sensitivity of $g_{a\gamma}\gtrsim2\times10^{-11}\,\mathrm{GeV}^{-1}$ using the heterodyne detection scheme~\cite{Hallal:2020ibe}.

For axion interferometry employing a photon counter, the expected photon rate in s~polarization is~\cite{2024PhRvD.109i5042Y}\footnote{
Inspecting the units of Eq.~(13) in Ref.~\cite{2024PhRvD.109i5042Y}, 
there should be a factor of $c^3$ instead of $c^5$.
}
\begin{equation}
    n_\mathrm{interf} = 4\frac{\hbar c^3}{\omega_a^2} G^2 g_{a\gamma}^2 \rho_a \epsilon_d \epsilon_a\frac{P_\mathrm{pump}}{h\nu},
\end{equation}
where $\omega_a = m_ac^2 / \hbar$ is the dark-matter axion frequency,  $\rho_a \approx 0.3\,\mathrm{GeV}\,\mathrm{cm}^3$ is the local dark matter density, and $P_\mathrm{pump}$ is the power of the pump laser in P~polarization. 
The cavity gain is $G = 2\sqrt{T_2}/(\sqrt{(1-T_1)(1-T_2)} - 1)$, where $T_1$ and $T_2$ are the transmissivities of the cavity mirrors with respect to the incident optical power. 
We assume values as for the LIDA cavity, that is, $T_1=17\,$ppm, $T_2 = 0.13\,\%$, $P_\mathrm{pump} = 118\,$kW~\cite{2024PhRvL.132s1002H}.
Again inserting $n_0$ from Eq.~\eqref{eq:flux_thr} for $n_\mathrm{interf}$ and solving for $g_{a\gamma}$ we find
\begin{equation}
    g_{a\gamma}^\mathrm{LIDA} \gtrsim 2.57^{+0.35}_{-0.30}\times10^{-14}\,\mathrm{GeV}^{-1}, 
\end{equation}
for $m_a = 2\,$neV ($\omega_a \approx 3\,$MHz).

\section{Conclusions and Outlook}
\label{sec:conclusion}

We have characterized a two-channel TES sytem optimized for the detection of near-infrared photons in terms of its SDE and energy resolution.
Using an experimental setup with a 1055\,nm cw laser, we have determined the SDE of both sensors to be $\sim(86\pm1)\,\%$.
Given the absorption efficiency of the optical stack in which the sensors are integrated, we expect a similar SDE for wavelengths between $\sim900$\,nm and $\sim1300\,$nm.
Both detectors show excellent energy resolution at 1055\,nm with $\Delta E / E  = (6.2\pm0.1)\,\%$ and $\Delta E / E  = (6.6\pm0.1)\,\%$ corresponding to FWHM values of $(0.172\pm0.002)\,$eV and $(0.183\pm0.001)\,$eV, respectively.

While we do not reach the high energy resolution of $3.5\,\%$ (at 0.8\,eV) reported in Ref.~\cite{2022SuScT..35i5002H}, our values are better than typical values of  $\sim10\,\%$~\cite[see, e.g.,][]{2022SuScT..35i5002H,2013ApPhL.102w1117L,2019ITAS...2909978K}

Using two lasers at different wavelengths, we have determined the energy calibration and developed a technique to select photon-like events based on the likelihood of a  multivariate normal distribution of pulse parameters determined from fits to time traces recorded with our modules. 
This procedure has the advantage that a single threshold value can be tuned to select photon-like events.

Using our photon selection, we find a bDCR of $(2.4 \pm 0.2)\times 10^{-3}\,$Hz and $(5.2 \pm 0.3)\times 10^{-3}\,$Hz for the two sensors in data runs lasting for 19 and 20\,hours, respectively.
We have developed an unbinned likelihood framework to fit our TES data simultaneously in the energy and time domain. 
With our unbinned likelihood approach, a spectral fit to the background data reveals that they are compatible with blackbody photons coupling form the laboratory environment at $\sim295\,$K to the optical fiber when the low-energy cutoff due to curling of the optical fiber is taken into account.
The main difference between the background spectra is found to be at low energies and probably due to slightly different fiber curling.
As the spectral fit with free fiber-curling parameters is sensitive to the initial choice of the fitting parameters, we plan to experimentally determine the low-energy cutoff in the future.  
We observe a slight excess of events over the blackbody background around $1.17\,$eV, which however is not found to be significant ($1.2\,\sigma$ and $2.5\,\sigma$ for TES1 and TES2, respectively). 

In absence of this potential additional background source, we have derived the sensitivity for our modules to a spectral line at $1.17\,$eV as expected in searches for hypothetical WISPs with LSW experiments and axion interferometers. 
Thanks to the energy resolution of our sensors, we find that a spectral lines at 1064\,nm with fluxes $\gtrsim3\times10^{-5}$\,Hz can be detected at $5\,\sigma$ significance within 20\,days.
Consequently, if our detector was integrated in a future ALPS~II science run, it should be almost compatible with the targeted  ALPS~II sensitivity of a photon-WISP coupling of $2\times10^{-11}\,\mathrm{GeV}^{-1}$.
In a LIDA-like axion interferometer, our TES should be sensitive to photon-WISP couplings of the order of $\sim3\times10^{-14}\,\mathrm{GeV}^{-1}$ at an WISP mass of $2\,$neV provided that the pump field can be suppressed sufficiently. 

We are investigating several improvements for different aspects of the work presented here. 
In the future, we plan to employ a pulsed laser source for SDE measurements to simplify the photon counting and statistical analysis~\cite{Schmidt2018}. 
The energy calibration and 
determination of the energy resolution 
will be improved through increasing the number of tested wavelengths, in particular at longer wavelengths. 
Furthermore, we are studying the performance of TES chips deposited on Si$_3$N$_4$ membranes which have been shown to have superior energy collection efficiencies and consequently better energy resolution~\cite{2008JLTP..151..125L}. 
The energy resolution can also be improved by fabricating TES chips with a lower critical temperature, as demonstrated in Ref.~\cite{2022SuScT..35i5002H}.
This should help to improve the sensitivity for spectral lines on top of a BBR background. 
Another possible avenue is the suppression of the blackbody background with the help of a cryogenic optical filter bench operated within the dilution refrigerator~\cite[e.g.,][]{Shibata2015}. 
An auto-alignable filter bench is currently under development.
Another possibility are narrow-band filters realized as fiber coating, which recently achieved $\sim10\,$nm transmission windows at 75\% transmissivity~\cite{2021SPIE11889E..0IB}. 

In terms of the data analysis, we plan to study more general distributions to describe the distribution of pulse parameters to improve the selection of photon-like events.
In this way, one could incorporate the energy dependence of the pulse parameters observed for TES2.
Instead of the derived pulse parameters, another avenue is to work with the raw time traces and (unsupervised) machine-learning algorithms to find hyperparameters distinguishing photon-like pulses from other background events (see, e.g., the discussion in Ref.~\cite{Rivasto:2025gjk}).
Lastly, a full background model also incorporating non-photonic events could be implemented in our unbinned likelihood framework. 
In this way, it would be possible to directly incorporate the likelihoods $\ln\pazocal{L}_\gamma$ describing how photon-like an event is in our statistical framework.

To summarize, we have shown that our tested module based on two tungsten TESs provided by NIST shows high quantum efficiency and excellent energy resolution.
Within our unbinned analysis framework, we have demonstrated the potential of these detectors for rare-event searches at optical and infrared wavelengths. 
Even without additional improvements on their design and background suppression, these cryogenic single-photon detectors should be able to detect low fluxes at current and future generation LSW experiments or axion interferometers.

\begin{acknowledgments}

We thank Adriana Lita from the National Institute of Standards and Technology, USA, for the TES devices, and Joern Beyer from Physikalisch-Technische Bundesanstalt, Berlin, Germany, for the SQUID modules, vital
advice and support. 
We also thank Marco Schmidt for his crucial guidance and assistance and Sebastian Raupach for helpful discussions on the experimental setup for measuring the system detection efficiency.
We thank our ALPS collaborators, 
in particular Aaron Spector for helpful discussions on the manuscript and Daniel Brotherton for helping with the spectrometer calibration. 
MM and ER
acknowledge the European Research Council (ERC) support under the European Union’s Horizon 2020 research and innovation program Grant agreement No. 948689 (AxionDM).
MM also acknowledges support from the Independent Research Fund Denmark (Grant No. 10.46540/5281-00207B).
We acknowledge the support by the Deutsche Forschungsgemeinschaft (DFG, German Research Foundation) under Germany’s Excellence Strategy – EXC 2121 “Quantum Universe” - 390833306 and by the Partnership between the University of Hamburg and DESY (PIER) - PIF-2023-16.
\end{acknowledgments}

\bibliography{main}

\appendix
\section{Calibration factors for photo diodes}
\label{sec:calpd}

Both PDs have two possible gain settings of ${\sim10^9}$\,V/W and ${\sim10^{10}}$\,V/W, which we refer to in the following as low gain and high gain, respectively.
The measured photocurrent is converted into the incident optical power through the calibration factors $m^g_\mathrm{cal,X}$ for the photodiodes X$=1, 2$,
\begin{equation}
    P_x = \frac{\bar{V}_x - \bar{V}_\mathrm{noise}}{m^g_\mathrm{cal, X}}, 
    \label{eq:v2p}
\end{equation}
where $\bar{V}_\mathrm{noise}$ is the average dark current, $g = (\mathrm{low}, \mathrm{high})$ is the gain for the used PD, and $\bar{V}_x$ is the mean of the voltage reading for measurement $x$.
The voltage is measured with an oscilloscope, connected to the PDs via low-noise coaxial cables.

Table~\ref{tab:pd-gain} shows the calibration factors for the photo diodes to convert the output voltage into incident optical power. 
The factors are derived from a linear fit (with intercept 0) to calibration data using a fiber-coupled laser source calibrated against known standards. 

\begin{table}[htb]
    \centering
    \caption{Calibration factors for both PDs and both gain settings.}
    \begin{tabular}{c|cc}
\hline
\hline
PD    & $m^\mathrm{low}_\mathrm{cal, X}$ ($10^{8}$ V / W) & $m^\mathrm{high}_\mathrm{cal, X}$ ($10^{8}$ V / W)\\
\hline
1   & $4.897 \pm 0.007$ &  $48.787 \pm 0.083$  \\
2   & $5.862 \pm 0.009$ &  $58.957 \pm 0.100$  \\
\hline
    \end{tabular}
    \label{tab:pd-gain}
\end{table}

\section{Details on Calibration and Derivation of Attenuation Factor}
\label{app:atten}

The total attenuation $A_\mathrm{tot}$ of the attenuation stage (see Fig.~\ref{fig:eff-setup}) is given by the product of the insertion losses of the MSs, $A_{\mathrm{MS}\,i}$, and the attenuation of the two VOAs, $A_\mathrm{VOA1}$ and $A_\mathrm{VOA2}$, which depend on the applied voltages, $V_1$ and $V_2$,
\begin{equation}
A_\mathrm{tot}(V_1, V_2) = \left(\prod_i A_{\mathrm{MS}\,i}\right) A_\mathrm{VOA1}(V_1)  A_\mathrm{VOA2}(V_2).
\end{equation}
As PD2 is less sensitive than the TES, we can only measure $A_\mathrm{tot}$ indirectly by decreasing the voltages applied to VOA1 and/or VOA2. 
We perform in total three measurements $x = 0,1,2$ for different settings for $V_1$ and $V_2$. 
For $x=0$ we set $A_0 \equiv A_\mathrm{tot}(0\,\mathrm{V}, 0\,\mathrm{V})$ (only insertion losses), for $x=1$ we take $A_1\equiv A_\mathrm{tot}(3.7\,\mathrm{V}, 0\,\mathrm{V})$ (VOA1 attenuation and insertion losses), and for $x=2$ we set $A_2\equiv A_\mathrm{tot}(0\,\mathrm{V}, 3.7\,\mathrm{V})$ (VOA2 attenuation and insertion losses).
In this way, we can express $A_\mathrm{tot}(3.7\,\mathrm{V}, 3.7\,\mathrm{V}) = A_1 A_2 / A_0$.\footnote{
In our notation, the attenuation factors $A_x$ and $A_\mathrm{tot}$ are unitless. 
They are related to the attenuation $L_x$ in dB through $L_x = 10 \log_{10} A_x$. 
} 
In addition, we decrease the voltage applied to VOA0 such that the total attenuation of the pre-attenuation stage is decreased from $\sim70\,$dB to $\sim50\,$dB.

We can now relate each $A_x$ to observable quantities.
For each measurement $x$, we measure a reference power $P_{\mathrm{ref}_x}$ at PD1 and an attenuated power $P_{A_x}$ at PD2, which are related in the same way as $ P_\mathrm{TES,\,in}$ and $P_0$ in Eq.~\eqref{eq:p-tes-in},
\begin{equation}
P_{A_x} = \frac{r'}{r} \frac{A_\mathrm{MS\,B'}} {A_\mathrm{MS\,A}}A_{x} P_{\mathrm{ref}_x} ~\Rightarrow~ A_x = \frac{r}{r'}\frac{P_{A_x}}{P_{\mathrm{ref}_x}}\frac{A_\mathrm{MS\,A}}{A_\mathrm{MS\,B'}},
\end{equation} 
which allows us to express $A_\mathrm{tot}$ as 
\begin{equation}
A_\mathrm{tot} = \frac{r}{r'}\frac{A_\mathrm{MS\,A}}{A_\mathrm{MS\,B'}}\frac{P_{A_1}}{P_{\mathrm{ref}_ 1}}\frac{P_{A_2}}{P_{\mathrm{ref}_2}}\frac{P_{\mathrm{ref}_0}}{P_{A_0}}.
\label{eq:Atot}
\end{equation}
We have introduced $A_\mathrm{MS\,B'}$ here, which is again the insertion loss at MS\,B, but might be different from $A_\mathrm{MS\,B}$ as we have to break the connection at MS\,B  to connect the attenuation stage to PD2 instead of the TES. 
We assume that $A_\mathrm{MS\,B} \approx A_\mathrm{MS\,B'}$ and include the uncertainty due to dis- and reconnecting the fiber in our final results. 
In this way, we arrive at our final expression for the SDE given in Eq.~\eqref{eq:sde-final} by plugging in $P_\mathrm{TES,~in}$ from Eq.~\eqref{eq:p-tes-in} into the SDE [Eq.~\eqref{eq:sde-base}] and using the expression for $A_\mathrm{tot}$ from Eq.~\eqref{eq:Atot},
\begin{equation}
\eta = n_\mathrm{TES}\frac{hc}{\lambda} \frac{1}{P_0} \frac{P_{\mathrm{ref}_1}}{P_{A_1}}\frac{P_{\mathrm{ref}_2}}{P_{A_2}}\frac{P_{A_0}}{P_{\mathrm{ref}_0}} = n_\mathrm{TES}\frac{hc}{\lambda} \frac{1}{A P_0},
\end{equation}
which is independent of $r, r', A_\mathrm{MS\,A}$, and $A_\mathrm{MS\,B}$ and is only related to observable quantities. 
Note that in the second step, we have introduced the final attenuation term
\begin{equation}
A = \frac{P_{A_1}}{P_{\mathrm{ref}_1}} \frac{P_{A_2}}{P_{\mathrm{ref}_2}}\frac{P_{\mathrm{ref}_0}}{P_{A_0}}.
\label{eq:atten-final}
\end{equation}
\\

\section{Error budget and individual measurement results for the SDE}
\label{app:results-errors}

In this appendix, we provide details on the derivation of the overall uncertainty on the SDE, $u(\eta)$, and the error budget. 
We also report the results of the individual measurements of $n_\mathrm{TES}$, $P_0$, and the wavelength $\lambda$ of the laser in Tab.~\ref{tab:eff-measurements} and for the attenuation factors in Tab.~\ref{tab:atten-factor}.

\begin{table*}[]
    \centering
    \caption{Results of the individual SDE measurements for SDE-A and SDE-B. Provided are the measurement values with uncertainties and relative uncertainties for $n_\mathrm{TES}$ and $P_0$. We also report the $p$-values of the KS test that the measured time intervals between photon pulses follow the expected exponential distribution. To obtain the efficiency values for SDE-B-tight, one has to rescale the corresponding values of $\eta$ with the average attenuation factor SDE-B-tight, see Tab.~\ref{tab:atten-factor}.}
    \label{tab:eff-measurements}
    \resizebox{\textwidth}{!}{%
    \begin{tabular}{c|cccccccccc}
\hline
\hline
Sensor & $\eta$ & $u(\eta)/\eta$ (\%) & $n_\mathrm{TES}$~(Hz) & $u(n_\mathrm{TES}) / n_\mathrm{TES}$ (\%) & $p_\mathrm{KS}$ & $P_0$ (pW) & $u(P_0) / P_0$ (\%) & $\lambda$ (nm) & $u(\lambda)/\lambda$ (\%)\\
\hline
\multicolumn{10}{c}{SDE-A}\\
\hline
TES1 &  0.773 $\pm$ 0.013 & 1.65 & 944 $\pm$ 15.4 & 1.63 & 0.115 & 73.50 $\pm$ 0.13 & 0.17 & 1055.0 $\pm$ 0.6 & 0.1  \\
TES1 &  0.852 $\pm$ 0.013 & 1.58 & 1039 $\pm$ 16.1 & 1.55 & 0.742 & 73.40 $\pm$ 0.13 & 0.17 & 1055.0 $\pm$ 0.6 & 0.1  \\
TES1 &  0.874 $\pm$ 0.014 & 1.56 & 1060 $\pm$ 16.3 & 1.54 & 0.517 & 73.02 $\pm$ 0.12 & 0.17 & 1055.0 $\pm$ 0.6 & 0.1  \\
TES1 &  0.872 $\pm$ 0.014 & 1.56 & 1057 $\pm$ 16.3 & 1.54 & 0.462 & 73.05 $\pm$ 0.12 & 0.17 & 1055.0 $\pm$ 0.6 & 0.1  \\
TES1 &  0.885 $\pm$ 0.014 & 1.55 & 1080 $\pm$ 16.4 & 1.52 & 0.776 & 73.42 $\pm$ 0.13 & 0.17 & 1055.0 $\pm$ 0.6 & 0.1  \\
\hline
TES2 &  0.872 $\pm$ 0.014 & 1.56 & 1059 $\pm$ 16.3 & 1.54 & 0.470 & 73.15 $\pm$ 0.13 & 0.17 & 1055.0 $\pm$ 0.6 & 0.1  \\
TES2 &  0.877 $\pm$ 0.014 & 1.56 & 1068 $\pm$ 16.3 & 1.53 & 0.571 & 73.32 $\pm$ 0.13 & 0.17 & 1055.0 $\pm$ 0.6 & 0.1  \\
TES2 &  0.889 $\pm$ 0.014 & 1.54 & 1086 $\pm$ 16.5 & 1.52 & 0.920 & 73.60 $\pm$ 0.13 & 0.17 & 1055.0 $\pm$ 0.6 & 0.1  \\
TES2 &  0.872 $\pm$ 0.014 & 1.56 & 1066 $\pm$ 16.3 & 1.53 & 0.522 & 73.59 $\pm$ 0.13 & 0.17 & 1055.0 $\pm$ 0.6 & 0.1  \\
TES2 &  0.849 $\pm$ 0.013 & 1.58 & 1038 $\pm$ 16.1 & 1.55 & 0.327 & 73.66 $\pm$ 0.13 & 0.17 & 1055.0 $\pm$ 0.6 & 0.1  \\
\hline
\hline
\multicolumn{10}{c}{SDE-B}\\
\hline
TES1 &  0.860 $\pm$ 0.014 & 1.59 & 1062 $\pm$ 16.3 & 1.53 & 0.503 & 76.76 $\pm$ 0.13 & 0.17 & 1056.7 $\pm$ 0.8 & 0.1  \\
TES1 &  0.863 $\pm$ 0.014 & 1.59 & 1063 $\pm$ 16.3 & 1.53 & 0.484 & 76.62 $\pm$ 0.13 & 0.17 & 1056.7 $\pm$ 0.8 & 0.1  \\
TES1 &  0.889 $\pm$ 0.014 & 1.57 & 1088 $\pm$ 16.5 & 1.52 & 0.995 & 76.10 $\pm$ 0.13 & 0.17 & 1056.7 $\pm$ 0.8 & 0.1  \\
TES1 &  0.854 $\pm$ 0.014 & 1.60 & 1046 $\pm$ 16.2 & 1.55 & 0.075 & 76.15 $\pm$ 0.13 & 0.17 & 1056.7 $\pm$ 0.8 & 0.1  \\
TES1 &  0.890 $\pm$ 0.014 & 1.57 & 1090 $\pm$ 16.5 & 1.51 & 0.201 & 76.17 $\pm$ 0.13 & 0.17 & 1056.7 $\pm$ 0.8 & 0.1  \\
\hline
TES2 &  0.818 $\pm$ 0.013 & 1.64 & 1000 $\pm$ 15.8 & 1.58 & 0.481 & 76.07 $\pm$ 0.13 & 0.17 & 1056.7 $\pm$ 0.8 & 0.1  \\
TES2 &  0.871 $\pm$ 0.013 & 1.53 & 1152 $\pm$ 17.0 & 1.47 & 0.847 & 82.26 $\pm$ 0.14 & 0.17 & 1056.7 $\pm$ 0.8 & 0.1  \\
TES2 &  0.862 $\pm$ 0.013 & 1.54 & 1141 $\pm$ 16.9 & 1.48 & 0.927 & 82.33 $\pm$ 0.14 & 0.17 & 1056.7 $\pm$ 0.8 & 0.1  \\
TES2 &  0.859 $\pm$ 0.013 & 1.54 & 1142 $\pm$ 16.9 & 1.48 & 0.792 & 82.65 $\pm$ 0.14 & 0.17 & 1056.7 $\pm$ 0.8 & 0.1  \\
TES2 &  0.838 $\pm$ 0.013 & 1.56 & 1109 $\pm$ 16.7 & 1.50 & 0.750 & 82.36 $\pm$ 0.14 & 0.17 & 1056.7 $\pm$ 0.8 & 0.1  \\
\hline
\end{tabular}
}
\end{table*}

Following \cite{Gerrits:2019idb}, we derive the relative uncertainty $u(\eta) / \eta$ on the SDE by propagation of errors, where we neglect any correlations between $P_0$ and $n_\mathrm{TES}$.
Starting from the expression of the SDE in Eq.~\eqref{eq:sde-final} with the average attenuation factor $\langle A \rangle$, the relative uncertainty is found to be 
\begin{widetext}
\begin{equation}
    \frac{u(\eta)}{\eta} = \sqrt{
    \left(\frac{u(n_\mathrm{TES})}{n_\mathrm{TES}}\right)^2
    + \left(\frac{u(\lambda)}{\lambda}\right)^2
    + \left(\frac{u(P_0)}{P_0}\right)^2
    + \left(\frac{u(\langle A \rangle)}{\langle A \rangle}\right)^2
    }.\label{eq:unc-sde}
\end{equation}
\end{widetext}
With $N$ counted photon pulses over the observation time $T$, the uncertainty on the photon rate is  
\begin{equation}
    u(n_\mathrm{TES}) = \frac{\sqrt{N}}{T} = \sqrt{\frac{n_\mathrm{TES}}{T}}.
\end{equation}
For the uncertainty on the wavelength, $u(\lambda)$, we use the FWHM of a Lorentzian profile fitted to the laser spectrum measured with one of the spectrometers. 
The two spectrometers were calibrated against an Nd:YAG crystal based non-planar ring oscillator (Coherent Mephisto 500) with a nominal wavelength of $(1064\pm0.5)$\,nm.
The HP4Pro spectrometer measured a central wavelength compatible with this value. However, the NIRQuest spectrometer was offset by $+3.4$\,nm with respect to the HP4Pro value and 
we have corrected for this offset in all subsequent measurements.
The measured optical power $P_0$ (and all other optical powers) is derived from the voltage readings of the calibrated PDs, see Eq.~\eqref{eq:v2p}. 
The uncertainty on the optical powers has thus three contributions, 
\begin{equation}
    \frac{u(P_x)}{P_x} = \sqrt{\left(\frac{u(m^g_\mathrm{cal, X})}{m^g_\mathrm{cal, X}}\right)^2
    + \left(\frac{u(\bar{V}_x)}{\bar{V}_x}\right)^2
    + \left(\frac{u(\bar{V}_\mathrm{noise})}{\bar{V}_\mathrm{noise}}\right)^2,
    }
\end{equation}
where $x$ refers to the power measurement in question. 
The uncertainties on the calibration factor  $m^g_\mathrm{cal, X}$ are derived from the fit to the calibration data and are reported in Appendix~\ref{sec:calpd}, and the uncertainties on the average voltage measurements are 
$u(\bar{V}_i) = s_{V_i} / \sqrt{N_{V_i}}$, with $s_{V_i}$ the standard deviation of the measured voltage trace and $N_{V_i}$ the numbers of samples in the trace. 
The uncertainties on $n_\mathrm{TES}$, $P_0$, and $\lambda$ of the individual measurements are also reported in Tab.~\ref{tab:eff-measurements}.

\begin{table}[htb]

    \caption{Individual and average attenuation factors derived from the measurements of optical powers with different attenuator settings, see Eq.~\eqref{eq:atten-final}.}
    \label{tab:atten-factor}

    \centering
    \begin{tabular}{cc}
\hline
\hline
$A~(\times 10^{-6})$ & $u(A) / A$ (\%) \\
\hline
\multicolumn{2}{c}{SDE-A}\\
\hline
 3.126 $\pm$ 0.012 & 0.381 \\
 3.111 $\pm$ 0.012 & 0.381 \\
 3.128 $\pm$ 0.012 & 0.381 \\
 3.138 $\pm$ 0.012 & 0.381 \\
 3.130 $\pm$ 0.012 & 0.381 \\

\hline
\multicolumn{2}{c}{$\langle A \rangle = (3.127 \pm 0.007)\times10^{-6}$}\\
\hline
\multicolumn{2}{c}{SDE-B}\\
\hline
 2.991 $\pm$ 0.011 & 0.381 \\
 3.029 $\pm$ 0.012 & 0.381 \\
 3.023 $\pm$ 0.012 & 0.381 \\
 3.046 $\pm$ 0.012 & 0.381 \\

\hline
\multicolumn{2}{c}{$\langle A \rangle = (3.022 \pm 0.011)\times10^{-6}$}\\
\hline
\multicolumn{2}{c}{SDE-B-tight}\\
\hline
 2.991 $\pm$ 0.011 & 0.381 \\
 2.985 $\pm$ 0.011 & 0.381 \\
 3.029 $\pm$ 0.012 & 0.381 \\
 3.023 $\pm$ 0.012 & 0.381 \\
 3.046 $\pm$ 0.012 & 0.381 \\

\hline
\multicolumn{2}{c}{$\langle A \rangle = (3.015 \pm 0.012)\times10^{-6}$}\\
\hline
    \end{tabular}
\end{table}

The attenuation factor $A$ of an individual measurement is given by Eq.~\eqref{eq:atten-final}. 
Using propagation of errors, the relative uncertainty is 
\begin{equation}
    \frac{u(A)}{A} =
    \sqrt{
    \sum_{x}\left(
    \frac{u(P_x)}{P_x}
    \right)^2
    }.
\end{equation}
The uncertainty of the average $\langle A \rangle$ is then found by adding the uncertainty $u(A)$ in quadrature with the uncertainty of the mean of $A$ given by the standard deviation $s_A / \sqrt{N_A}$, where $N_A$ is the number of individual measurements of $A$.
The individual uncertainties and uncertainty of the averages are reported in Tab.~\ref{tab:atten-factor}.

In the same way, we derive the final uncertainties on the average SDE values reported in Tab.~\ref{tab:eff-measurements}, i.e., we sum in quadrature the uncertainties of the individual measurements from Eq.~\eqref{eq:unc-sde} with the standard deviation of the SDE values divided by the square root of  the number of measurements, $s_\eta / \sqrt{N_\eta}$.

\end{document}